\documentclass[sigconf]{acmart}
\renewcommand\footnotetextcopyrightpermission[1]{} 
\usepackage{graphicx} % 图片包
\usepackage{hyperref} % Ensure hyperref is loaded first
\usepackage{hyperxmp}
\usepackage{subcaption} % 子图和子标题包
\usepackage{algorithm}
\usepackage{algorithmic}
\usepackage{tabularx, booktabs}
\usepackage{graphicx}
\usepackage{multirow}
\usepackage{ulem}
\usepackage{setspace}
\usepackage{xcolor}
\usepackage{array}
\usepackage{balance}
\usepackage{enumitem}
\usepackage{booktabs}
\usepackage{ulem}
\usepackage{CJKutf8}
\usepackage{makecell}
\usepackage{geometry}%页面设置
\usepackage{subcaption}
\usepackage{stfloats}
\usepackage{graphicx} % 图片包
\usepackage{subcaption} % 子图和子标题包
\usepackage{siunitx}
\usepackage{amsmath}
\usepackage{amssymb}
\usepackage{float}
\usepackage{acronym}

\acrodef{IR}{Information Retrieval}
\acrodef{MLP}{multilayer perceptron}
\acrodef{SERP}{Search Engine Result Page}
\acrodef{ERP}{event related potential}
\acrodef{EEG}{electroencephalogram}
\acrodef{DT}{Gradient Boosting Decision Tree}
\acrodef{SST}{SST-EmotionNet}
\acrodef{AUC}{Area Under Curve}
\acrodef{IN}{Information Need}
\acrodef{fMRI}{functional magnetic resonance imaging}
\acrodef{BCI}{brain–computer interface}
\acrodef{BP}{band power}
\acrodef{DE}{differential entropy}
\acrodef{RASM}{rational asymmetry}
\acrodef{DASM}{differential asymmetry}
\acrodef{SVM}{support vector machines}
\acrodef{DBN}{deep belief networks}
\acrodef{KNN}{k-Nearest Neighbors}
\acrodef{HBG}{Height-Biased Gain}
\acrodef{OP}{Opinion Polarization}

\AtBeginDocument{%
  \providecommand\BibTeX{{%
    \normalfont B\kern-0.5em{\scshape i\kern-0.25em b}\kern-0.8em\TeX}}}

\makeatletter
\def\hlinew#1{%
  \noalign{\ifnum0=`}\fi\hrule \@height #1 \futurelet
   \reserved@a\@xhline}
\makeatother

\setcopyright{acmcopyright}
\copyrightyear{2021}
\acmYear{2021}
\acmDOI{10.1145/1122445.1122456}

\acmConference[SIGIR 2025]{SIGIR 2025: THE 48TH INTERNATIONAL ACM SIGIR CONFERENCE ON RESEARCH AND DEVELOPMENT IN INFORMATION RETRIEVAL}{JULY 13-18, 2025}{PADUA, ITALY}
\begin{document}
% \fancyhead{}

% 要不要加Relevant？
\title{Understanding the Effect of Opinion Polarization in Short Video Browsing}


\author{Bangde Du}
\email{dbd23@mails.tsinghua.edu.cn}
\affiliation{%
  \institution{BNRist, DCST, Tsinghua University}
  \country{China}
}

\author{Ziyi Ye}
\email{yeziyi1998@gmail.com}
\affiliation{%
  \institution{BNRist, DCST, Tsinghua University}
  \country{China}
}

\author{Monika Jankowska}
\email{mj46@rice.edu}
\affiliation{%
  \institution{Rice University}
  \country{USA}
}

\author{Zhijing Wu}
\email{wuzhijing.joyce@gmail.com}
\affiliation{%
  \institution{BNRist, DCST, Tsinghua University}
  \country{China}
}

\author{Qingyao Ai}
\email{aiqy@tsinghua.edu.cn}
\affiliation{%
  \institution{BNRist, DCST, Tsinghua University}
  \country{China}
}

\author{Yiqun Liu}
\email{yiqunliu@tsinghua.edu.cn}
\affiliation{%
  \institution{BNRist, DCST, Tsinghua University}
  \country{China}
}

\renewcommand{\shortauthors}{Du, et al.}

\begin{CJK}{UTF8}{gbsn}
\begin{abstract}

This paper explores the impact of Opinion Polarization (OP) in the increasingly prevalent context of short video browsing, a dominant medium in the contemporary digital landscape with significant influence on public opinion and social dynamics.
We investigate the effects of OP on user perceptions and behaviors in short video consumption, and find that traditional user feedback signals, such as like and browsing duration, are not suitable for detecting and measuring OP. 
Recognizing this problem, our study employs Electroencephalogram (EEG) signals as a novel, noninvasive approach to assess the neural processing of perception and cognition related to OP. 
Our user study reveals that OP notably affects users' sentiments, resulting in measurable changes in brain signals.
Furthermore, we demonstrate the potential of using EEG signals to predict users' exposure to polarized short video content. 
By exploring the relationships between OP, brain signals, and user behavior, our research offers a novel perspective in understanding the dynamics of short video browsing and proposes an innovative method for quantifying the impact of OP in this context.

\end{abstract}
\end{CJK}
\settopmatter{printacmref=false}

\begin{CCSXML}
<ccs2012>
   <concept>
       <concept_id>10002951.10003317.10003331</concept_id>
       <concept_desc>Information systems~Users and interactive retrieval</concept_desc>
       <concept_significance>500</concept_significance>
       </concept>
   <concept>
       <concept_id>10002951.10003317</concept_id>
       <concept_desc>Information systems~Information retrieval</concept_desc>
       <concept_significance>500</concept_significance>
       </concept>
    <concept>
    <concept_id>10003120.10003130.10011762</concept_id>
    <concept_desc>Human-centered computing~Empirical studies in collaborative and social computing</concept_desc>
    <concept_significance>500</concept_significance>
    </concept>
 </ccs2012>
\end{CCSXML}

\ccsdesc[500]{Information systems~Information retrieval}
\ccsdesc[500]{Information systems~Users and interactive retrieval}
\ccsdesc[500]{Human-centered computing~Empirical studies in collaborative and social computing}

\keywords{Opinion Polarization, Short Video Browsing, EEG}


%%
%% This command processes the author affiliation and title
\maketitle
\begin{CJK}{UTF8}{gbsn}
\section{Introduction}
In recent years, internet-based social platforms have become increasingly important for the dissemination of information and the formation of public opinion.
However, these platforms have also been criticized for their role in promoting Opinion Polarization~\cite{pariser2011filter,hosanagar2014will} across many aspects, including politics~\cite{groshek2017helping,martinovic2018exploring}, healthcare~\cite{holone2016filter}, and science~\cite{mccright2011politicization}.
\ac{OP} refers to the widening divergence of individuals' stances on specific issues, where stances shift from moderate, centrist opinions to more extreme opinions.
OP is now recognized as a significant factor in propagating or even creating biases that can influence decisions and opinions~\cite{pariser2011filter,abebe2018opinion}.
For example, \citet{gao2023echo} revealed that users on short video platforms often encounter content tailored to specific polarized opinions, leading to the development of entrenched and emotionally charged sentiments toward political issues.
\citet{asker2019thinking} observed a significant correlation between user's emotional intensity and OP in social media.

% Introduce existing solutions to measure and alleviate OP: using co-click behaviors to similar items measure OP, e.g. click entropy. 
% (can discard: Alleviate OP by several solutions, e.g., balancing the ratio of exploration and exploitation in recommender systems.) 
As the phenomenon of \ac{OP} becomes more prevalent in the digital age, existing efforts have been undertaken to understand the phenomenon of \ac{OP} by quantifying its strength and mitigating its effect on societal polarization.
Typically, they detect and quantify OP through artificially designed metrics based on clicks or other behavior signals collected in the logs of internet-based social platforms. 
For instance, some studies starting from an individual perspective argue that greater click entropy on content with similar opinions indicates stronger OP~\cite{chen2022more}. 
Others, approaching from a group perspective, cluster users based on their behavior signals on the same items and define the strength of OP according to the distance between various clustering groups~\cite{chitra2020analyzing}.
Based on these OP measurements, they proposed several methods to alleviate the potential 
negative impact of OP, 
including more exploratory recommendation systems~\cite{chen2021values,chen2021exploration}, 
presenting opposing view~\cite{einav2022bursting},
and group discussion~\cite{strandberg2019discussions}. 
% Introduce the weakness of existing research: co-click signals are solely implicit feedback from users, which is biased, and not accurate. 
% (Showing some evidence here) 
% There is no guarantee that these implicit user behaviors can reflect the strength of OP. Whether OP has effect on user perceptions of short video content remains unknown. 
% Here, perception means sentiment and affective state. Address RQ1.
The above studies are limited to using user behavior metrics as a quantification of OP.
% On the one hand, these behavior signals are solely implicit feedback from users, which might be biased, and not accurate~\cite{ye2022don,wang2021clicks,joachims2017accurately}.
% For example, existing studies have demonstrated that clicks do not necessarily indicate a user's agreement with content~\cite{lu2018between}, and non-clicks do not invariably imply disinterest in an item presented in social platforms~\cite{williams2016detecting, ye2022don}. 
Unfortunately, there is no guarantee that these user behaviors, especially implicit user behaviors, can directly reflect the effect of OP on human users. 
For example, \citet{criss2021twitter} observed that on Twitter~\footnote{\url{https://twitter.com/}}, users' click and like behavior on some polarized content does not necessarily alter their opinions or attitudes towards the polarization of certain content.
Hence, existing methods for detecting and quantifying \ac{OP} may be fraught with inaccuracies and fail to directly reflect the impact of OP on users.
How to understand and quantify OP in terms of its impacts on users in social media content is still an open question to the research communities. 
% Moreover, it has not been conclusively proven whether exposure to content with polarized information influences users' pre-existing views and attitudes.~

% 显示观点记录
To mitigate the issue of solely relying on user behavior metrics, some existing qualitative research has delved into the phenomenon of OP by collecting several explicit responses, such as interviewing~\cite{mas2013differentiation} and analysis on content posted by individual users~\cite{lee2016impact}. 
Their research has demonstrated that OP can affect an individual's sentiment~\cite{mas2013differentiation,criss2021twitter}, affective state~\cite{lee2016impact}, and several other complex cognitive processes regarding a specific topic in social platforms~\cite{druckman2021affective}.
However, collecting explicit signals requires a lot of user effort and there is a lack of systematic effort in connecting explicit signals with OP in a quantitive manner.
% it is not intuitive to transform these signals into quantifiable metrics.
Few studies have considered how to utilize these explicit responses to detect and quantify OP, particularly in the context of internet-based social platforms such as short video streaming platforms and personalized search engines.

Recently, researchers have attempted to user's brain signals as a measurement to understand several concepts in information access and interaction systems, such as relevance judgment~\cite{ye2023relevance,moshfeghi2013understanding}, information need~\cite{moshfeghi2019towards}, knowledge state~\cite{pinkosova2023moderating}, etc.
As a portable and economical device for collecting brain signals, \ac{EEG} has been a widely used neurological measurement for understanding users' perception, attention, memory, and affective state during information processing.
%As a quantitative measure of the users’ neural processing of perception and cognition in real-time, \ac{EEG} provides a robust and reliable alternative to self-reporting without additional user effort.
\ac{EEG} can provide a quantitative measure of the users' neural processing of perception and cognition in real-time, and it can be collected during the task process without interfering with the user. 
Therefore, we believe that \ac{EEG} could complement other behavioral signals and provide important insights for understanding the connections between short video browsing and opinion polarization.
% In IR, they are also widely applied as a measure to evaluate information accessing performance. 
Hence, EEG is also a potential alternative to explicit response and behavior signals for understanding user's neural processing of content perception in OP scenarios. 

In this paper, we aim to predict and quantify the occurrence of OP, and explore the relationship between OP and various signals including user's behavior signals, explicit responses, and brain signals, respectively.
Specifically, we focus on the following research questions:
\begin{itemize}
    \item \textbf{RQ1:} How does opinion polarization in short video browsing affect users' sentiment judgment？
    \item \textbf{RQ2:} How do opinion polarization exposure affect users' brain signal patterns?
    \item \textbf{RQ3:} How can we predict the possibility of users' exposure to short videos with polarized opinions? 
\end{itemize}

%\todo{I suggest using personages or figures instead of characters throughout the paper.}
To shed light on these research questions, we conduct a user study examining the influence of browsing polarized short videos about different historical personages on human participants.
We conduct a multi-faceted analysis with various user signals collected: sentiment annotations, behavior signals, and EEG signals collected during the short video browsing process.
Based on the user study, we reveal that OP has a significant effect on users' sentiment judgments of the personages.
In addition, we also observe a detectable difference between users' brain signal response to the video content of these personages before and after the influence of OP.
These differences in brain activity can be underpinned by various neuroscientific factors such as emotions, memory, and attention during the user's perception procedures.
Finally, we conduct a classification experiment using behavior signals, explicit annotations, and brain signals, as well as their possible combinations to predict a participant's exposure to polarized short videos.
The experimental results suggest that quantification and detection of OP based on explicit responses and brain signals are more accurate compared to previous methods that relied on implicit behavior signals.

\section{Methodology}
In this section, we first present how we selected target personages, constructed a video set, and designed a video platform for the experiment, as the preliminary part. Then we introduce the procedure of the whole experiment, followed by the details about participants and ethic-related information. The dataset can be found in the link below~\footnote{The data and code are available at %\url{https://anonymous.4open.science/r/EEG_and_Explicit_Feedback_for_Opinion_Polarization_Detection-0E18}}.
\url{https://github.com/bangdedadi/EEG-VideoPolarization}}

\begin{figure*}[h]
  \centering
  \includegraphics[width=\textwidth]{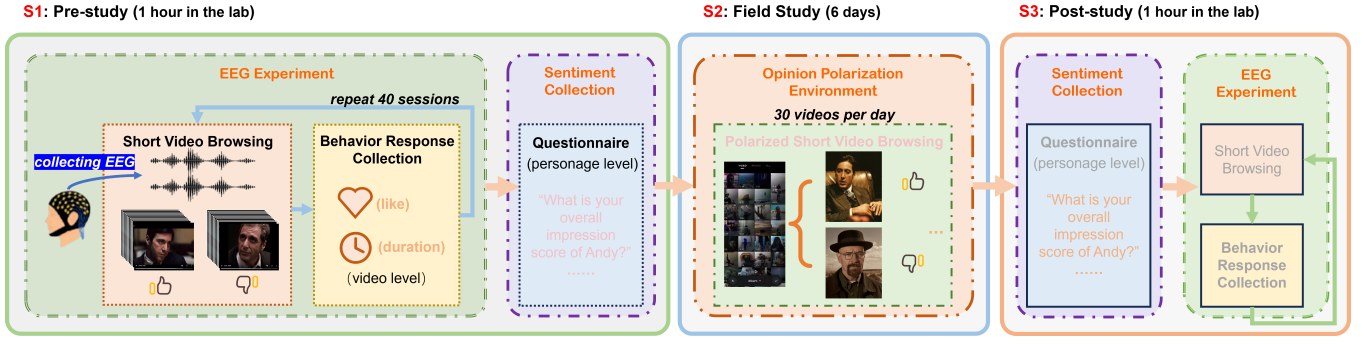}
  \caption{The overall procedure of the three-stage user studies.
  %\todo{use different video screenshots; explicitly annotate $S_i$ as the $i^{th}$ stage.} 
  $S_1$: Participants browsed videos featuring positive or negative polarity on these 10 personages. Their behavior signals, brain signals, and explicit responses are collected during the video browsing process or the post-experiment questionnaire. $S_2$: Participants browsed short videos about these personages posted by the Platform in a field study lasting 6 days. For each personage, the Platform only recommended videos with either positive or negative polarity, which acts as a manifestation of OP.
  $S_3$: The last stage resembles the procedures in stage $S_1$. However, it diverges in the selected videos and the timing of the questionnaire.
}
  \label{fig:precedure}
\end{figure*}

%\subsection{Task preparation}
\subsection{Preliminary}
\subsubsection{Target personages selection}
%To study how short video browsing affects people's opinions, we selected ten historical or fictional personages based on the following criteria: (1) the personages should be famous or controversial enough so that we could find short videos with different opinions on them; (2) the personages should not be so popular that participants in our experiments have already developed strong opinions on the personages by their own.
We selected ten historical or fictional personages based on two criteria: (1) sufficient fame or controversy to ensure polarized video content, and (2) moderate popularity to avoid pre-existing strong opinions among participants.
Specifically, based on the background of the participants (discussed in Section 2.3), we selected personages including Thomas Alva Edison, Cao Cao, Christopher Columbus, Emperor Wu of Han, Catherine the Great, Walter White, Michael Corleone, Isaac Newton, Qin Shi Huang, and Xiang Yu.
The introductions to these personages are shown in Table~\ref{intro_of_personage}.
We investigated the familiarity of the participants with these personages through questionnaires in the pre-study and post-study.
According to the questionnaires, before our experiments, participants had a moderate level of familiarity, averaging a score of 2.8 on a scale from 1 to 5. After our experiments, this average familiarity score increased to 3.2.

\begin{table}[t]
\centering
\caption{Brief introductions of the historical and fictional personages selected in our experiment.}
\begin{tabular}{>{\small\bfseries}p{1.5cm} p{6.3cm}}
\toprule
\multicolumn{1}{c}{\textbf{Name}} & \multicolumn{1}{c}{\textbf{Introduction}} \\
\midrule
Thomas Alva Edison & A prolific American inventor known for the light bulb, phonograph, and motion picture camera, impacting modern life. \\ \hline
Cao Cao & A prominent warlord and statesman at the end of the Eastern Han dynasty in ancient China, known for his military and political acumen. \\ \hline
Christopher Columbus & An Italian explorer in the 15th century, credited with the discovery of the Americas while searching for a new route to Asia. \\ \hline
Emperor Wu of Han & The seventh emperor of the Han dynasty in China, known for his military conquests and the expansion of the Chinese empire. \\\hline
Catherine the Great & Empress of Russia in the late 18th century, renowned for her expansion of the Russian empire and domestic reforms. \\\hline
Walter White & A fictional character from the TV series "Breaking Bad," a high school chemistry teacher who turns to cooking methamphetamine. \\ \hline
Michael Corleone & A fictional character in "The Godfather" movie series, transforms from a reluctant outsider to a ruthless mafia boss. \\\hline
Isaac Newton & A key figure in the scientific revolution, an English mathematician, physicist, and astronomer, known for his laws of motion and gravity. \\\hline
Qin Shi Huang & The founder of the Qin dynasty and the first emperor of a unified China, famous for the Terracotta Army and the Great Wall. \\\hline
Xiang Yu & A prominent military leader and political figure in ancient China during the late Qin dynasty, known for his role in the Chu–Han Contention. \\
\bottomrule
\end{tabular}
\label{intro_of_personage}
\end{table}

\subsubsection{Video set construction}
To conduct experiments on short video browsing, We constructed a video set regarding the ten personages for the user study. The video set construction involved video collection and annotation processes. Using keyword searches, we collected videos themed around the target personages from the Tiktok~\footnote{\url{https://www.douyin.com/}} and bilibili~\footnote{\url{https://www.bilibili.com/}}. 
The principles for selecting videos are (1) the main topic of the video should focus on the target personage; and (2) the video should exhibit clear polarity (i.e., positive, negative). To avoid potential effects caused by video length, each video was truncated to a duration of one minute without affecting the key content of the video. In total, we collected 16 videos for each personage.

% To categorize the videos into positive and negative polarities, we annotated them accordingly.
The videos were categorized into positive and negative polarities by human annotation.
Three expert annotators who had extensive experience in video opinion annotations were recruited to annotate the polarity of each video toward the corresponding personage.
Each video is annotated ranging from 1 to 5 following a Likert scale, where 1 represents the most negative opinion, while 5 signifies the most positive opinion. 
The polarity of the videos was determined based on the average scores from three experts, where scores above and below 3 denote positive and negative polarity, respectively. 
The annotation guideline can be found in the dataset repository\footnote{The annotation guideline is available at %\url{https://anonymous.4open.science/r/EEG_and_Explicit_Feedback_for_Opinion_Polarization_Detection-0E18/README.md}}.
\url{https://github.com/bangdedadi/EEG-VideoPolarization/blob/main/README.md}}

To better simulate short video browsing scenarios and prevent the participants from identifying the goal of our lab and field studies, we also included a background video set in our experiments. This set contains 120 one-minute clips unrelated to the target personages, each also edited to one minute in duration. Additionally, 40 videos from the SEED-IV dataset (a well-known dataset for sentiment analysis)\cite{zheng2018emotionmeter},
%used as controls and references for Valence and Arousal annotations,
were included and truncated to a one-minute duration.
The final video set comprised three types of videos: \textbf{positive videos about target personages (pos-videos), negative videos about target personages (neg-videos), and clips unrelated to the personages (distractors).} 

\subsubsection{Platform}

\begin{figure}[t]
  \centering
  \includegraphics[width=0.9\linewidth]{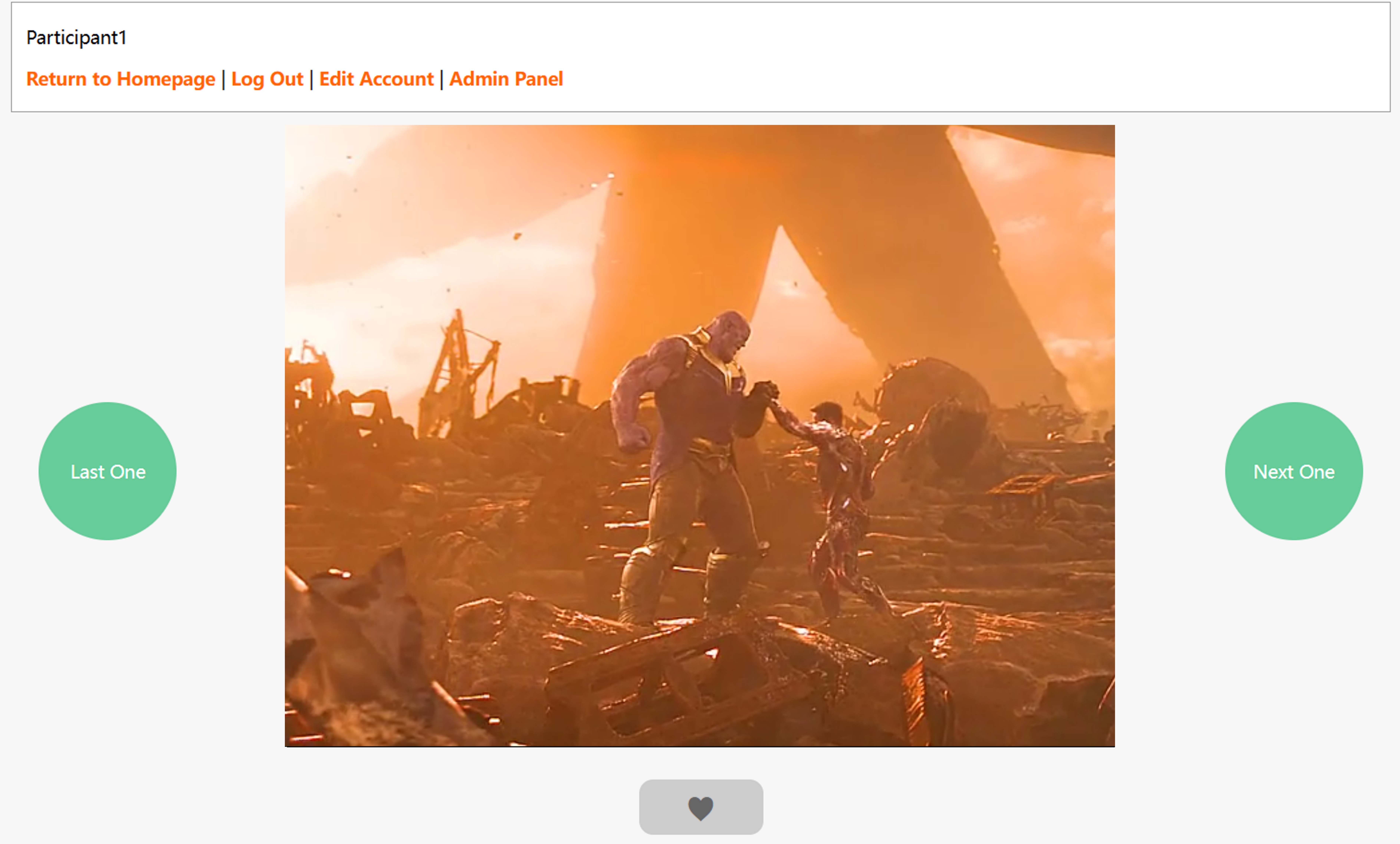}
  \caption{Webpage platform playback page.}
  \label{fig:webpage}
  \vspace{-10pt} % 调整下面的空格
\end{figure}

To facilitate data collection and mitigate the influence of habitual use of a single platform, a custom experimental webpage for video playback was developed for this study. 
The webpage can collect user interactions, including ``likes'' and viewing duration, mimicking standard short video platforms. 
To approximate real-world conditions, the webpage is designed for both mobile phones and laptops. 
As shown in Figure~\ref{fig:webpage}, the webpage's playback interface primarily features a video frame, a "Next One" button to skip to the next video, a "Last One" button to go back to the previous video, and a heart button for "likes". We provide unique accounts to each participant, and they have to log in before start browsing the video.

\subsection{Procedure}
The procedure of our user study is shown in Figure~\ref{fig:precedure}. 
Our experiment contains three stages: pre-study, field study, and post-study.

\textbf{Pre-study.}
The goal of the pre-study is to collect participants' initial opinions and their brain signal patterns when watching the video of different personages. Specifically, we collected the participants' brain signals with an EEG device. In the pre-study stage, participants were asked to first go through a brief training process to ensure that they were familiar with our short video browsing platform and understand the experiment procedure.
When the experiment officially started, participants were instructed to stay relaxed throughout the process and avoid large physical movements or frequent blinking to minimize potential noise interference with EEG signals. 

Each participant was required to watch 40 videos, each lasting one minute. 
These 40 videos included 20 videos about specific personages, which we refer to as the personage-type videos, and 20 extracted from SEED-IV~\cite{zheng2018emotionmeter} (a well-known dataset for sentiment analysis), which we refer to as the SEED videos.
The SEED videos serve as ``distractors" to prevent the participants from identifying the goal of our study at the beginning of our experiments, otherwise, they could manipulate the experiment results intentionally.
Breaks were included between video sessions, allowing participants to rest, maintain their attention, and minimize fatigue. 
Each personage-type video talks about one personage, and there were 10 personages involved in our experiment in total.
Specifically, in the pre-study stage, each personage was associated with two videos: one presenting a positive opinion and the other a negative opinion. 
Each video is 60 seconds in duration. 
The first 30 seconds are presented, after which users are allowed to skip the rest of the video by clicking the "Next One" button. 
All participants watched the same group of videos, but the order in which the videos were presented was randomized for each participant to prevent any systematic bias due to the order of presentation.

After the EEG experiment, as depicted in Figure~\ref{fig:precedure}, participants were instructed to complete a questionnaire. 
The questionnaire comprised 40 items, with 20 questions about the experiment personages. 
For each personage, participants were asked to (1) provide a sentiment score from 1 (poor impression) to 5 (excellent impression), expressed as "Intuitively, what is your overall impression score of [the personage]?"; (2) rate their familiarity from 1 (strongly disagree) to 5 (strongly agree) with each personage, phrased as "Do you consider yourself well-acquainted with [the personage]?".
The remaining 20 questions focus on the content of the background video and are unrelated to the selected experimental personages, with the same 1-5 scale. For example, ``How would you rate Jackie Chan's performance in `The Accidental Spy'?"(there is a background video on Jackie Chan and his movie "The Accidental Spy"). These distractors help us prevent the participants from identifying the objective of our experiment and manipulate their responses on purpose. Questions in the questionnaire are randomized to mitigate the potential order effect. All data relating to the questionnaires can be found in the dataset~\footnote{The data relating to the questionnaires is available at %\url{https://anonymous.4open.science/r/EEG_and_Explicit_Feedback_for_Opinion_Polarization_Detection-0E18/data/questionnaire/readme.md}}.
\url{https://github.com/bangdedadi/EEG-VideoPolarization/tree/main/data}}

\textbf{Field Study.}
To investigate the impact of opinion polarization (OP) on individuals, in the field study, we constructed two types of environment bias for each personage used in the pre-study stage: one fostering a positive opinion, and the other a negative opinion. 
For each participant, each of the personages shown to them was randomly assigned with one type of bias. Specifically, if a pair consisting of a personage and an experiment participant is assigned a positive or negative bias, all the videos related to the personage presented to the participant during the field study would portray a favorable/adverse stance towards that target personage. 

During the field study stage, participants were required to log in to a designated account to browse short videos every day. 
Each participant was required to watch 30 one-minute videos daily. 
Among these videos, 10 videos talked about 10 different target personages, each conveying either a positive or negative opinion about the respective personage. 
The remaining 20 videos were from the background distractors, aiming at creating a more realistic information environment and preventing the participants from focusing excessively on any single personage.
The sequence of these 30 videos was randomized for each participant. The likes and viewing duration for each video playback were recorded. 

To ensure participants watched the videos attentively, we sent a daily question about the video content.
The questions were straightforward and could be answered easily if the participants had watched the videos. For example, ``Have you watched any content related to `The Hulk'? Please describe the main plot". In our experiments, all participants correctly answered the daily questions. The Field Study began on the day following the completion of the pre-study and extended over six days.

\textbf{Post-study.}
The post-study consists of a questionnaire and an EEG experiment. Both steps resemble the procedures elaborated in the pre-study. The only difference between the pre-study and the post-study is the timing of the questionnaire. In the pre-study, questionnaires were given after the EEG experiments, while in the post-study, they were given before them. This ensured that any differences observed between the two questionnaires were attributable solely to the field study and not influenced by the laboratory studies.

\subsection{Participants and Ethics}
In total, our experiment recruited 24 participants, who are active users of short video platforms, with ages ranging from 18 to 25 years. Participants comprised 13 males and 11 females from various academic backgrounds, 
including Computer Science, Environmental Science, Automation, and Life Sciences. 
All participants are right-handed and reported no history of neurological disorders, ensuring uniformity in terms of brain function and manual dexterity for the tasks involved in the study. Each participant was remunerated approximately 50 USD for their involvement.
Note that, after the whole experiment procedure, we explicitly informed the participants about our experiment objectives and let them decide whether we could use their data for future analysis to avoid potential ethical risks.
Due to the limited budgets and the need for an in-person lab study for EEG data collection, we could only recruit local participants around our experiment location. 
Thus, our participants are not diverse in terms of race and culture, and the personages and videos we selected for the experiments are more or less tailored to the homogeneous background of our participants. 
We acknowledge that this is a limitation of our experiment and leave the study of this limitation to future study.

The collected dataset consists of 24 participants, among which one participant's data was discarded for technical issues, yielding data for 230 user-personage pairs. Each pair includes EEG signals, behavior responses (i.e., like and duration) while viewing the video about the specific personage, and sentiment annotations on the personage. 

%This study has been reviewed and approved by an Ethics Committee~\footnote{Anonymized during the double-blinded review process}.
This study has been reviewed and approved by the Department of Science and Technology Ethics Committee, Tsinghua University(THU01-20230221).
We ensured the experiments were harmless to all participants. Participants were required to sign a consent form before the user study. The consent form detailed the nature and purpose of the study and emphasized the confidentiality and anonymity of responses. It also assured participants of their right to withdraw from the study at any point without any penalty.

\section{Result Analysis}
To comprehensively analyze the impact of opinion polarization in short video contexts, we conducted an analysis in Section 3.1 on ``Explicit Feedback through Annotations'', which includes sentiment annotations (i.e., sentiment scores collected in the questionnaire) and behavioral responses (i.e., like rates and viewing duration ratios), and represented the findings in bar graphs. In Section 3.2, we focused on ``Implicit Feedback through Brain Signals''. We explored the connection between opinion polarization and brain activity.

\subsection{Questionnaire and External Behaviors}
To more comprehensively analyze the impact of opinion polarization, we examined sentiment annotations through the questionnaire and analyzed external user behaviors.

\subsubsection{Sentiment annotations on personages before and after the field study}
To examine the influence of Opinion Polarization (OP) on user sentiment, we collected sentiment scores for each personage from participants before and after the field study using a five-point Likert scale. Control sentiment data for distractors, unrelated to the field study entities, were also gathered.

\begin{figure}[t]
  \centering
  \begin{subfigure}{0.49\columnwidth}
    \includegraphics[width=\linewidth]{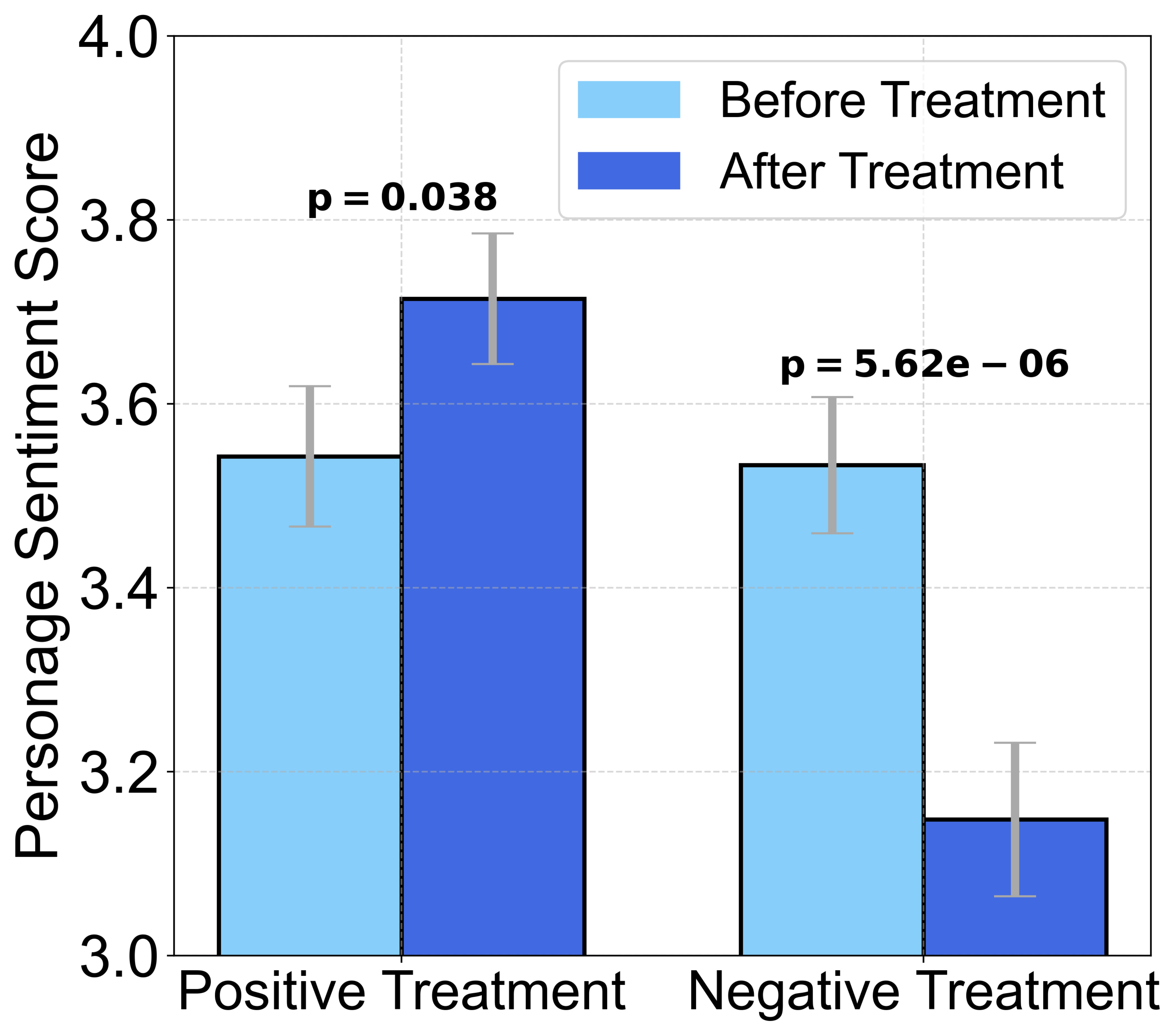}
    \caption{Sentiment for personages}
    \label{fig:sentiment_sub1}
  \end{subfigure}
  \hfill % 在两个子图之间添加一些空间
  \begin{subfigure}{0.49\columnwidth}
  \includegraphics[width=\linewidth]{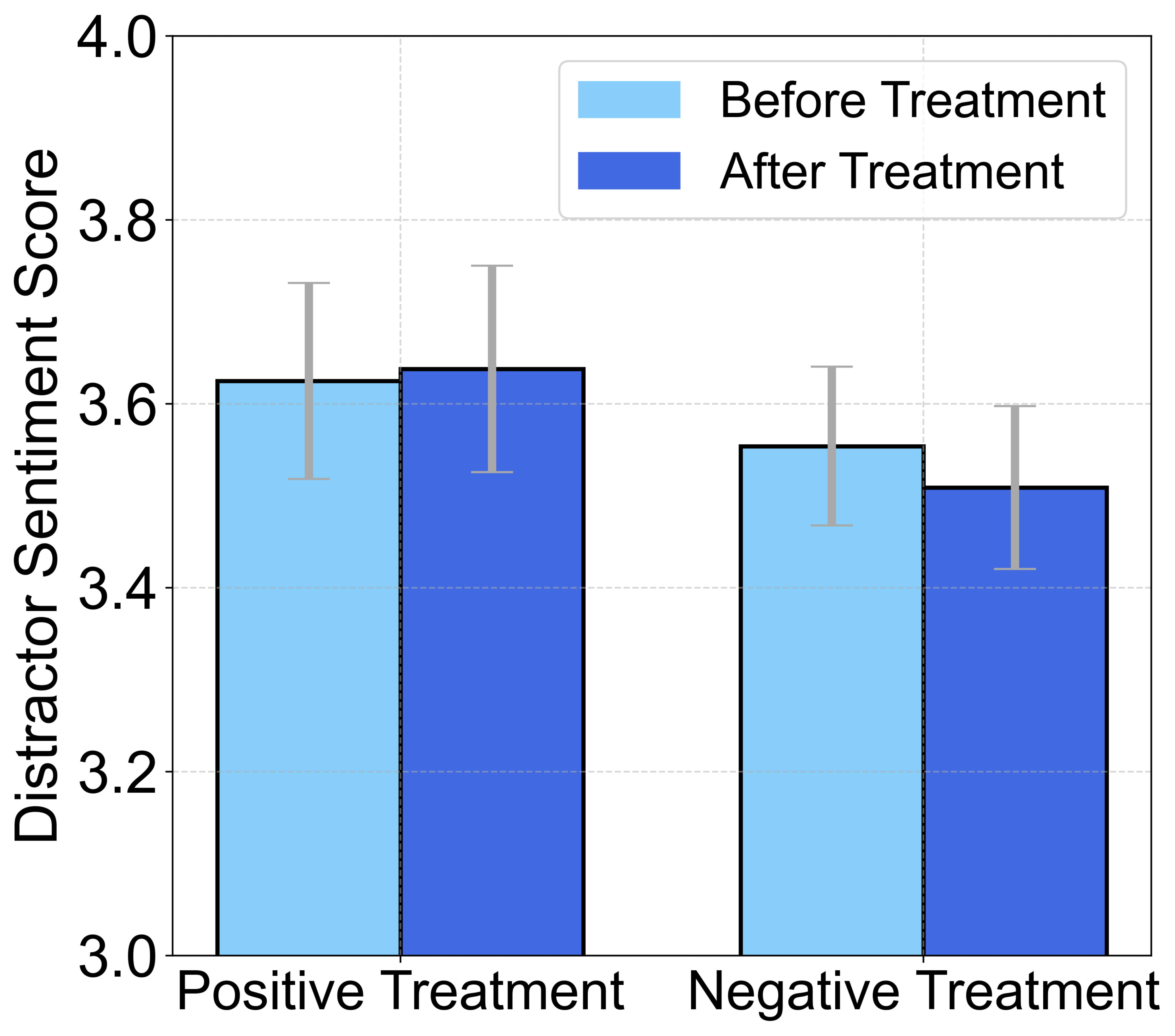}
  \caption{Sentiment for Distractors}
  \label{fig:sentiment_sub2}
  \end{subfigure}
\caption{Sentiment scores averaged across participants and personages in pre- and post-OP environments.}
\label{fig:sentiment scores}
\vspace{-2mm}
\end{figure}

We calculated the average sentiment scores for all personages in both pre- and post-OP environments, as shown in Figure~\ref{fig:sentiment scores}.  
The OP environments are either positive or negative.
Before exposure, there was no significant difference in average sentiment scores between the environments (3.543 and 3.533).
Sentiment scores increased significantly~(p=$0.038$) in the positive environment, from 3.543 (Pre Positive) to 3.714 (Post Positive). 
In the negative environment, scores decreased notably~(p=$5.62e^{-6}$) from 3.533 (Pre Negative) to 3.148 (Post Negative).
The sentiment scores for background videos (distractors) showed no significant change.
These observations indicate that the OP environment could significantly affect users' sentiment judgment towards personages.
%This impact stems from direct user feedback, not indirect analysis of platform logs.
%\todo{emphasis on pvalue}

\begin{figure*}[t]
  \centering
  \includegraphics[width=0.8\textwidth]{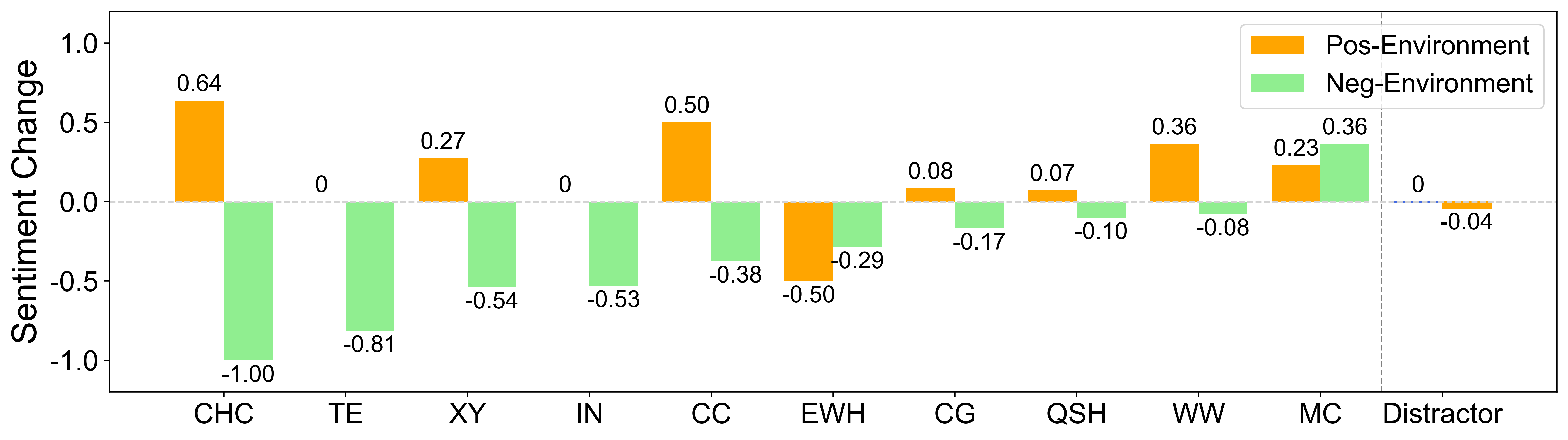}
  \caption{Difference of sentiment score collected pre- and post-environment for different personages and distractors, averaged across participants. CHC, TE, ..., and MC are the abbreviations of the ten selected personages. Distractor indicates the average score across all distractors for the experimental control.}
  \label{fig:everyperson}
\end{figure*}

To illustrate OP's impact, we computed the change in sentiment scores pre- and post-exposure for each personage.
We found \textbf{a significant difference (p=0.0153 < 0.05)} in sentiment changes between environments.
%Figure~\ref{fig:everyperson} displays the sentiment changes for ten personages, represented by abbreviations on the ten horizontal axis items on the left, after exposure to positive or negative opinion environments. 
As shown in Figure~\ref{fig:everyperson},  for personages like Christopher Columbus (CHC), Thomas Alva Edison (TE), and others, the positive environment caused a positive sentiment change, while the negative environment caused a negative change.
However, Emperor Wu of Han (EWH) and Michael Corleone (MC) exhibited different sentiment changes in positive and negative environments.
We analyzed three experts’ annotations on video opinions and calculated consistency scores (kappa scores) among personages, as shown in Table~\ref{tab:kappa}.
%Fleiss’ Kappa is a statistical measure used to assess the reliability of agreement among a fixed number of raters when they assign categorical ratings to several items or classified items. 
Fleiss’ Kappa measures the reliability of agreement among raters assigning categorical ratings.
Notably, Emperor Wu of Han (EWH) and Michael Corleone (MC) showed lower consistency, possibly explaining their anomalous sentiment changes. The lowest Kappa value for a personage reached the fair agreement level~\cite{landis1977measurement}. We presented experimental results for all personages to ensure completeness.
%\label{appendix:annotation guideline}

%Fleiss' Kappa is a statistical measure used to assess the reliability of agreement among a fixed number of raters when they assign categorical ratings to several items or classify items. Fleiss' Kappa scores calculated for each personage are shown in Table~\ref{tab:kappa}:

\begin{table}[t]
\centering
\caption{Fleiss' Kappa Values for Various Personages}
\small
\begin{tabular}{l|c}
\toprule
\textbf{English Name} & \textbf{Fleiss' Kappa} \\
\midrule
Thomas Alva Edison (TE) & 0.402 \\
Cao Cao (CC) & 0.570 \\
Christopher Columbus (CHC) & 0.728 \\
Emperor Wu of Han (EWH) & 0.251 \\
Catherine the Great (CG) & 0.553 \\
Walter White (WW) & 0.417 \\
Michael Corleone (MC) & 0.270 \\
Isaac Newton (IN) & 0.483 \\
Qin Shi Huang (QSH) & 0.383 \\
Xiang Yu (XY) & 0.321 \\
\bottomrule
\end{tabular}
\label{tab:kappa}
\end{table}
%Details of video annotation consistency scores for all personages are included in Appendix~\ref{appendix:annotation guideline}.

\subsubsection{Behavior Response during short video browsing}
We analyzed the collected behavioral data, including likes and viewing durations. 
The like rate and viewing duration ratio are calculated as follows:
\begin{equation}
\text{Like Rate} = \frac{\text{\# Liked videos}}{\text{\# Videos}}
\end{equation}

\begin{equation}
%\text{Viewing Duration Ratio} = \frac{\sum (\text{Viewing Duration of Video } T_i)}{\text{\# Videos} \times 60\, \text{s}}
\text{Viewing Duration Ratio} = \frac{\text{Total Viewing Time}}{\text{\# Videos} \times \text{Duration per Video}}
\end{equation}

\begin{figure}[t]
  \centering
  \begin{subfigure}{0.49\columnwidth}
    \includegraphics[width=\linewidth]{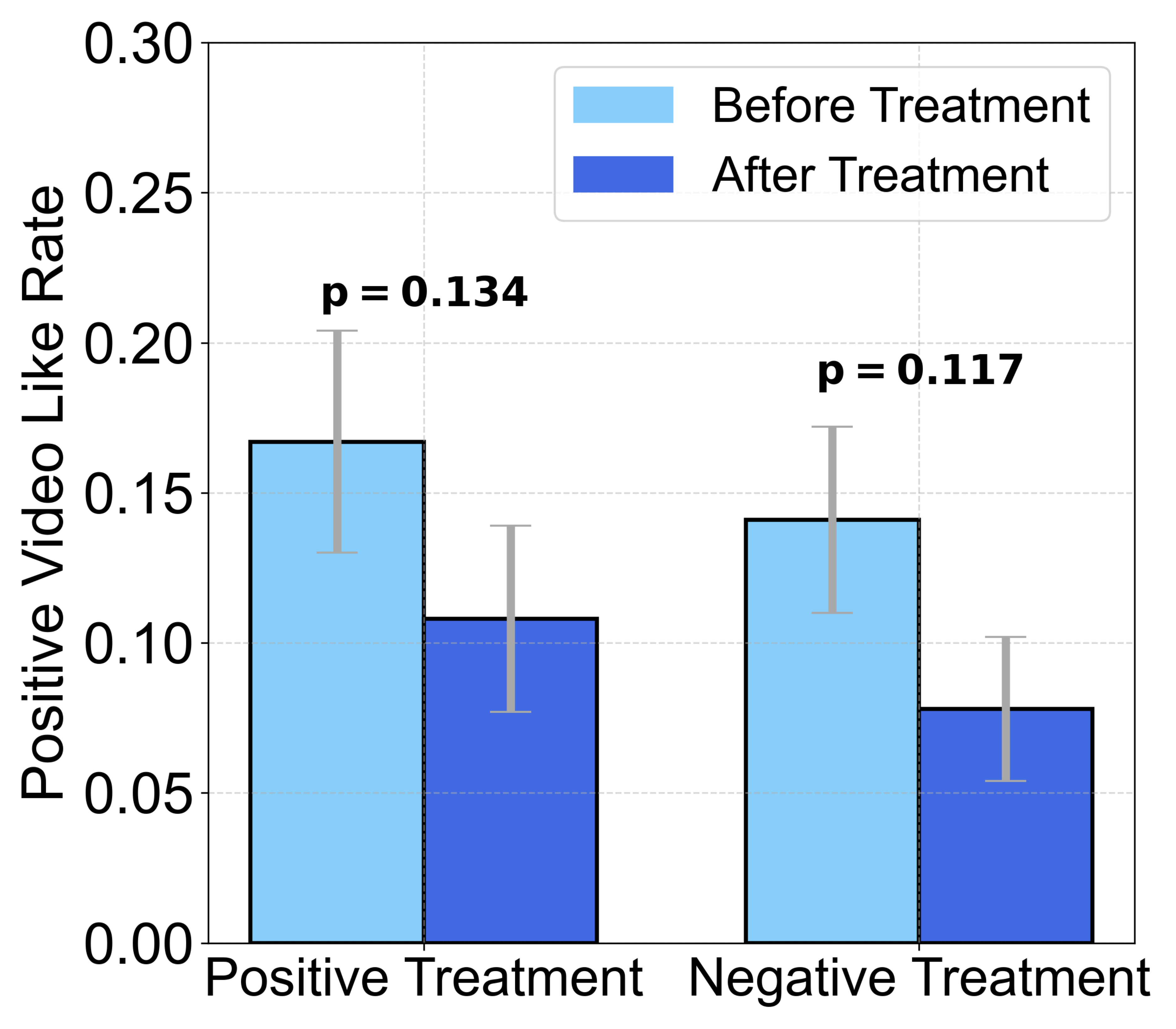}
    \caption{Pos-videos}
    \label{fig:like_sub1}
  \end{subfigure}
  \hfill % 在两个子图之间添加一些空间
  \begin{subfigure}{0.49\columnwidth}
  \includegraphics[width=\linewidth]{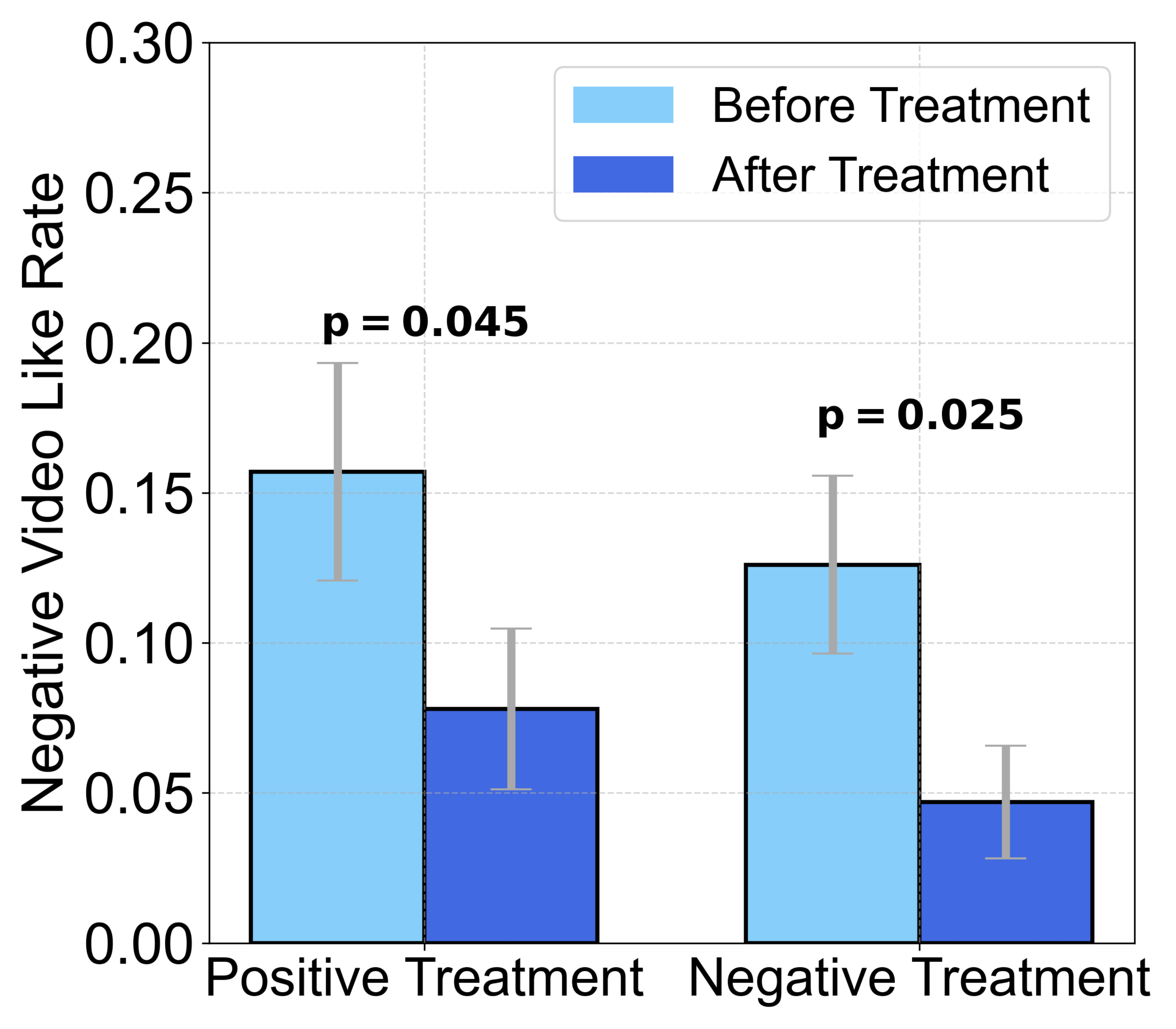}
  \caption{Neg-videos}
  \label{fig:like_sub2}
  \end{subfigure}
\caption{Averaged like rate for videos with positive~(Pos-videos) or negative opinions~(Neg-videos).}
\label{fig:likerate}
\vspace{-2mm}
\end{figure}

Figure~\ref{fig:likerate} illustrates a significant decline in the like rate when users, following exposure to either positive or negative opinion environments, watch videos with both positive and negative opinions. This indicates a diminished user interest in monothematic videos. A comparison of the two OP environments' impacts revealed no significant difference in the like rates between positive and negative influences. This might suggest that like data, traditionally used as a behavioral metric, are less effective in probing OP's impact on users

\begin{figure}[t]
  \centering
  \begin{subfigure}{0.49\columnwidth}
    \includegraphics[width=\linewidth]{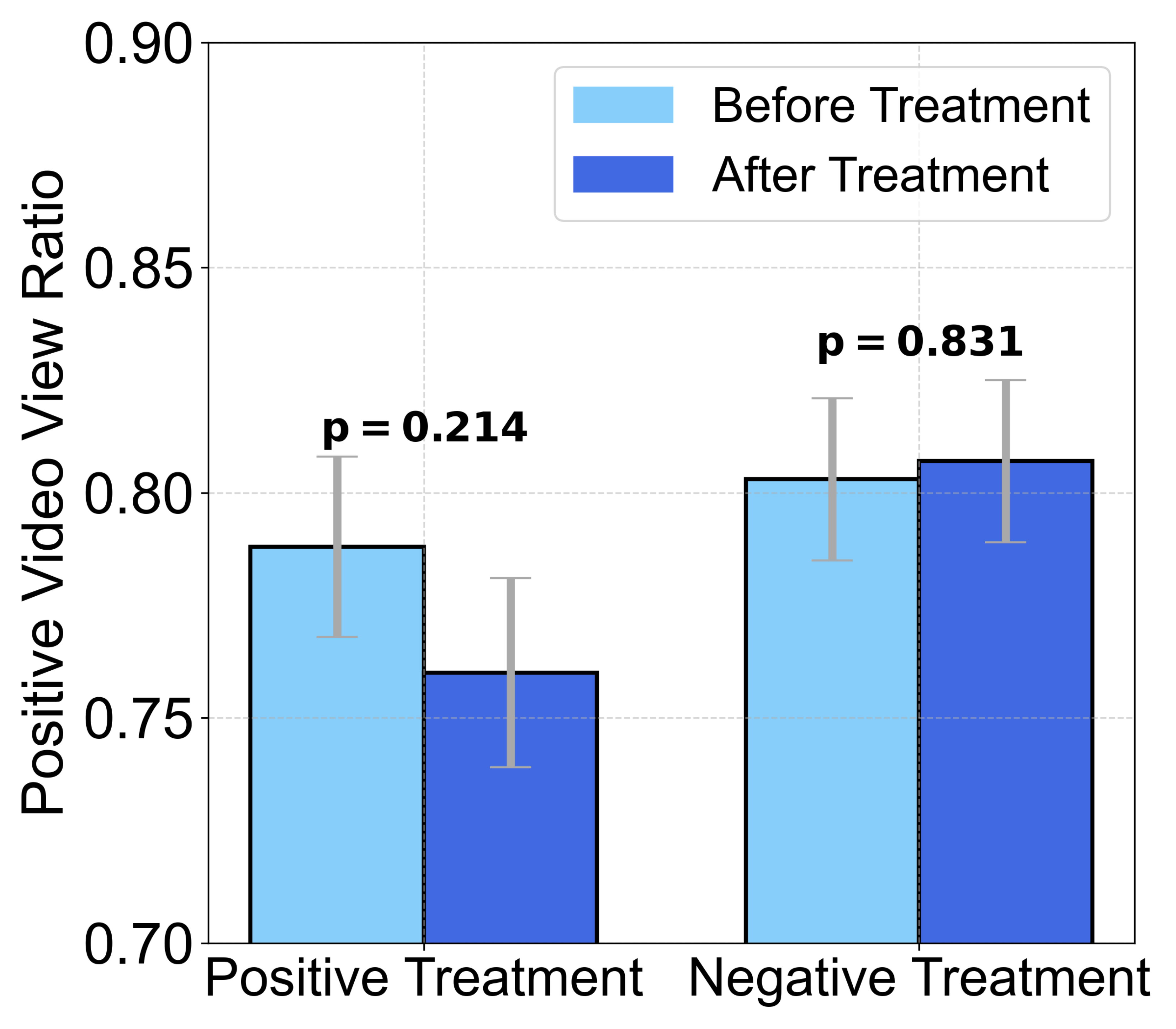}
    \caption{Pos-videos}
    \label{fig:time_sub1}
  \end{subfigure}
  \hfill % 在两个子图之间添加一些空间
  \begin{subfigure}{0.49\columnwidth}
  \includegraphics[width=\linewidth]{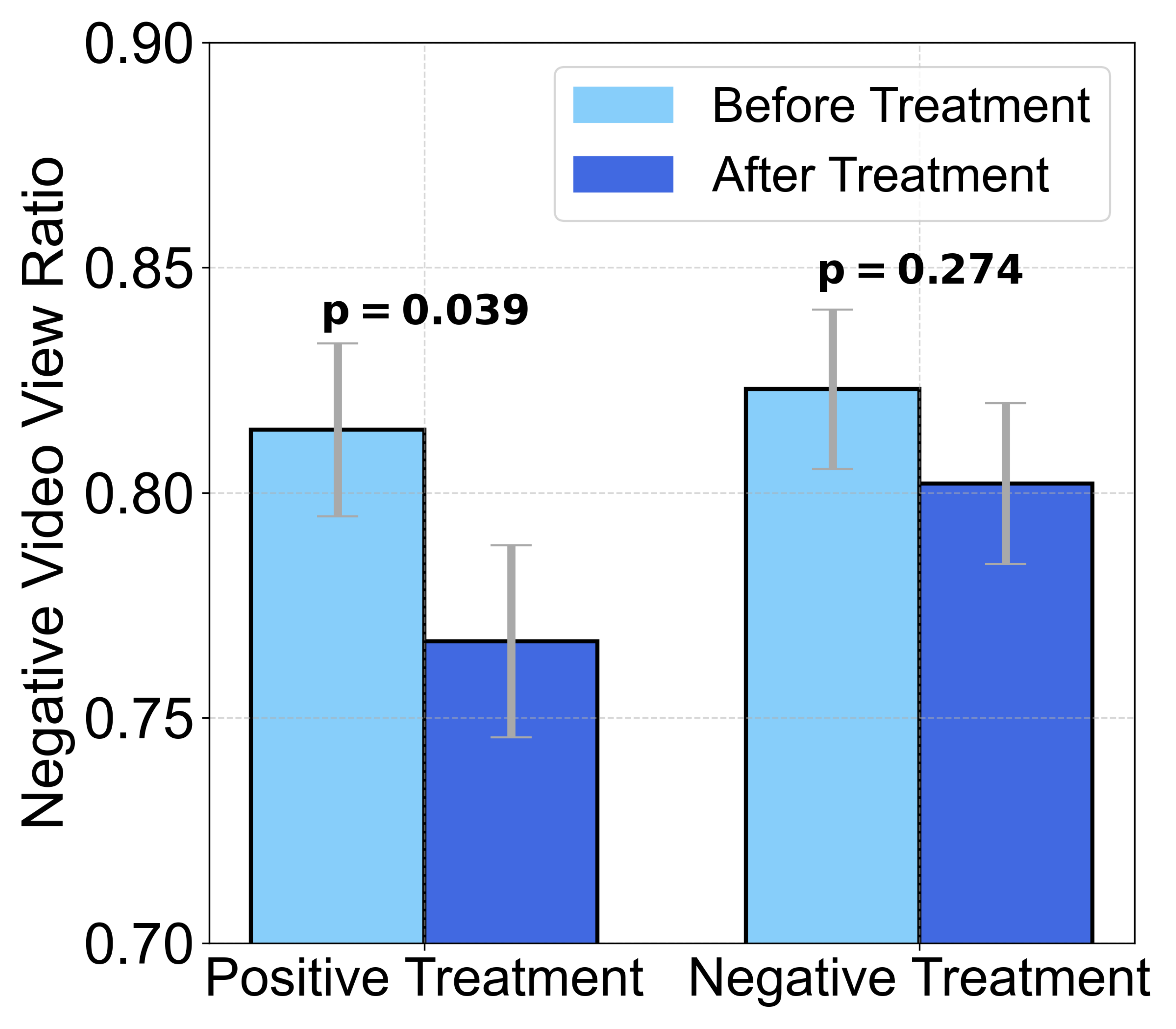}
  \caption{Neg-videos}
  \label{fig:time_sub2}
  \end{subfigure}
\caption{Averaged viewing duration ratio for videos with positive~(Pos-videos) and negative opinion~(Neg-videos). }
\label{fig:behavior}
\vspace{-2mm}
\end{figure}

As shown in Figure~\ref{fig:behavior}, the viewing duration ratio, averaged for positive and negative polarity videos, ranges from 0.5 to 1.0 minute, as the users are allowed to skip a video after 0.5 minutes and the total duration of each video is 1 minute. From Figure~\ref{fig:behavior}, we observe that post-exposure to a positive opinion environment, the viewing duration ratio for both positive and negative opinion videos marginally decreased. 
In summary, relative to sentiment score graphs, the viewing duration ratio did not exhibit a notable difference in the influence of positive versus negative opinions on users.

\paragraph{Answer to \textbf{RQ1}:} Sentiment score analysis reveals that opinion polarization significantly influences user sentiments, making them more aligned with the prevailing opinions of the environment. On the other hand, the user behavior data (like and viewing duration) has a relatively smaller effect associated with OP environments. This implies that explicit sentiment scores potentially offer a more accurate reflection of the impact of OP than implicit behavioral metrics.

\subsection{Brain Signals}

\begin{figure*}[t]
  \centering
  \includegraphics[width=0.8\textwidth]{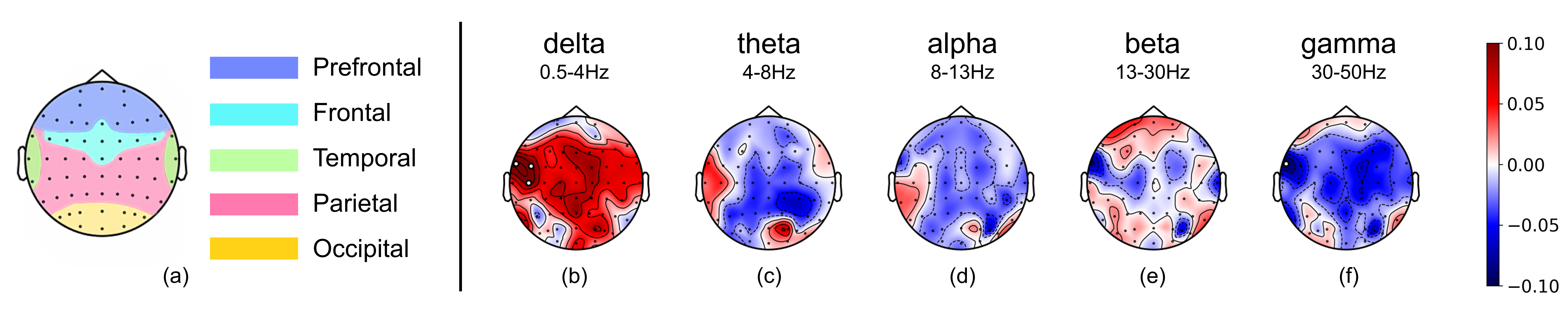}
  \caption{Pearson correlations between opinion polarization and differential entropy (DE) in EEG frequency bands: delta (0.5-4 Hz), theta (4-8 Hz), alpha (8-13 Hz), beta (13-30 Hz), and gamma (30-50 Hz) during the laboratory study. A white circle represents a significant correlation (p < 0.05) between DE features and OP environments. Panel (a) illustrates the electrode brain regions.}
  \label{fig:overallcor}
\end{figure*}

\subsubsection{Neural Correlates of Opinion Polarization in EEG}
To investigate the relationship between EEG signals and opinion polarization, we calculated the Pearson correlation coefficients between the Differential Entropy (DE) feature of EEG and the OP for each user-video pair. Specifically, we assessed the correlation between OP environments (positive or negative) in the field study and the EEG signals collected during the post-study. For each user-video pair, we averaged the DE feature across all frequency bands and electrodes and then computed the Pearson correlation coefficients~\cite{he2023understanding,schober2018correlation}. 

Figure~\ref{fig:overallcor}~(b-f) presents the correlations between OP environments and DE.
In the delta frequency band, a relatively strong positive correlation exists between positive OP environments and EEG signals, suggesting that positive environments may enhance low-level EEG activity in the Delta band. Conversely, negative OP environments are associated with a strong negative correlation in the Delta band with EEG signals. Existing research has highlighted an increase in delta wave activity in the frontal lobe during unconsciousness~\cite{magnus1987zeta}. This indicates that after experiencing positive OP environments, when users are exposed to videos of the same personage again, the unconscious part of their brain activity is relatively more engaged.% compared to after experiencing negative OP.

We observed that positive OP environments generally exhibit a strong negative correlation with EEG signals in the gamma band.
The gamma band has been recognized as crucial for learning and memory processes~\cite{headley2011gamma} and has also been correlated with meditation~\cite{kumar2021increased}. 
This suggests that after exposure to negative OP environments, when users view videos on the same personage again, the emotional response activities may be more active than exposure to positive OP environments.

\subsubsection{Neural Correlates of Opinion Polarization in EEG with behaviors}

\begin{figure}[t]
\centering
\includegraphics[width=\linewidth]{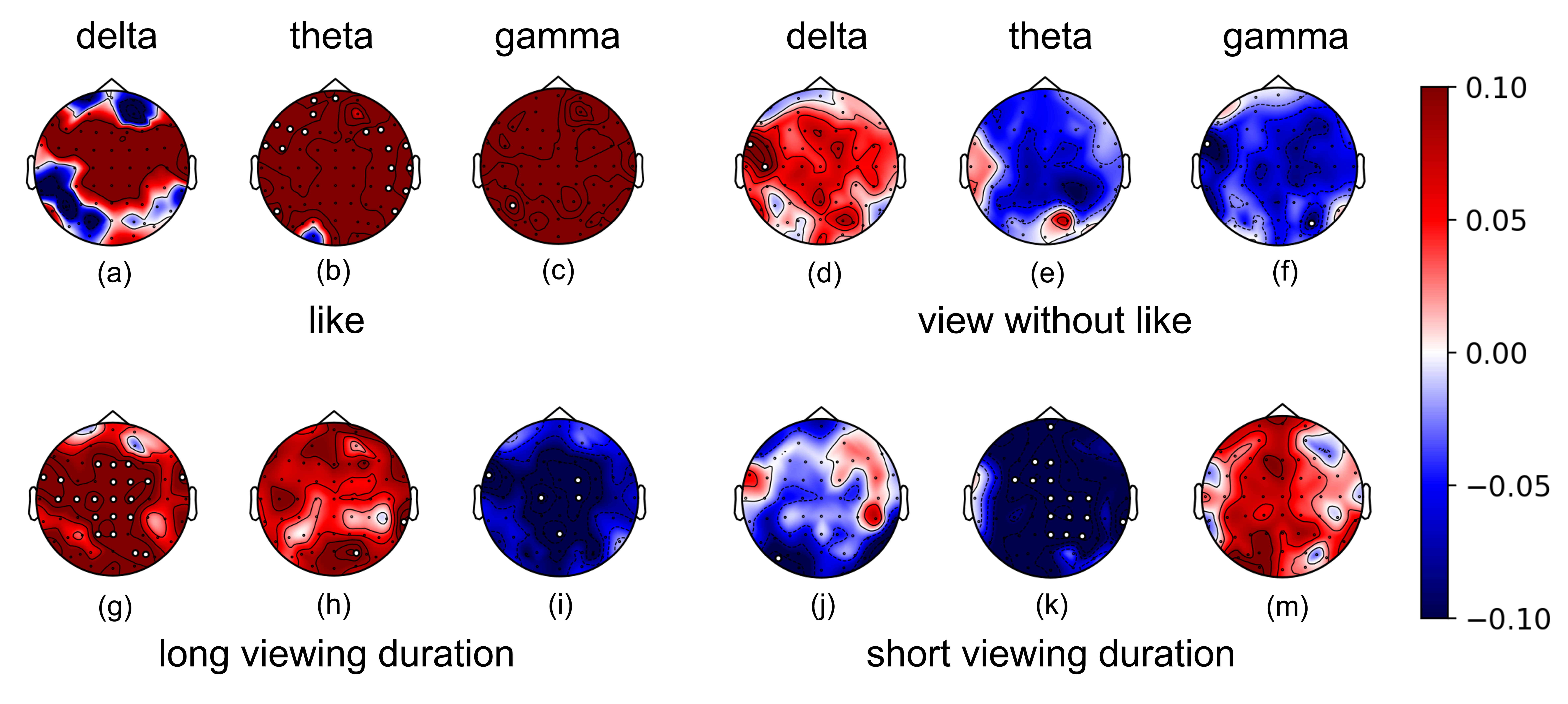}
\caption{The Pearson correlations of the opinion polarization with
DE in different frequency bands grouped by like data~(like or view without like, a-f) and viewing duration~(long or short, g-m).}
\label{eeg_behavior}
\vspace{-2mm}
\end{figure}

To explore the impact of behavioral data on the correlation between OP and EEG, brainwave signals were categorized based on 'likes' received and viewing duration. Videos were classified into two categories: those with a viewing duration ratio above 0.9 as long viewing duration, and those below 0.9 as short viewing duration. This method yielded two balanced data sets, with an approximate ratio of low to high viewing duration sets being 10:9.

Analysis of Figure~\ref{eeg_behavior}~(a-f) revealed a notable increase in the positive correlation between the different frequency bands of OP environments and EEG signals in the ``like'' group. 
This implies that for videos receiving likes in post-study, stronger EEG signals across multiple frequency bands are observed after positive OP environments. 
Significantly, correlations were found in the prefrontal and frontal lobe region of the theta band, an area linked to the generation and processing of emotions~\cite{ertl2013emotion}.

As shown in Figure~\ref{eeg_behavior}~(g-m), grouping the data by viewing duration yielded similarly significant results. 
In the long viewing duration group, a majority of electrodes in the delta band, mainly in frontal and temporal regions, showed a notable positive correlation, while the gamma band showed significant negative correlations. Long viewing durations suggest a higher willingness to watch~\cite{merrill2019go}. 
For short viewing durations, significant correlations were observed in the theta band, especially in the prefrontal and frontal regions.

\paragraph{Answer to \textbf{RQ2}:} 
EEG signals reveal significant links with OP, where Delta and Gamma bands indicate differing impacts of positive and negative polarization on brain activity.
Furthermore, we reveal that such links are dependent on users' behavioral response (i.e., `likes' and viewing duration). 
Videos receiving more likes or with longer viewing duration show stronger correlations, indicating that user engagement behavior in general plays a crucial role in detecting the impact of OP.

\section{Opinion polarization detection}

Motivated by the observed correlation between OP and user signals, we further investigate to what extent OP can be detected with those signals and address RQ3.
This section first elaborates on the task formalization and experiment setup of the proposed OP detection task, followed by studying the effect of various user signals, including sentiment judgment, behavior responses, and brain signals, in OP detection.

\subsection{Task Formalization}
\label{5.1}

To investigate ways to mitigate opinion polarization, especially in the context of specific topics or personages, it is essential to have a reliable method for determining the prevailing type of opinion polarization, be it positive or negative.
Consequently, we formalize the detection of OP as a binary classification problem. 
We utilize various user signals, including sentiment judgment, behavior responses, and brain signals as input, aiming to predict whether users experience a positive or negative polarity (i.e., the direction of OP) regarding a particular topic.

In each participant and personage pair~(e.g., participant A and personage Michael Corleone), we designed a setting to reflect positive opinion polarization. During the field study, Participant A was consistently shown videos that positively depicted Mike. For this specific pair, participant A's EEG and behavioral responses were recorded while watching Mike-themed videos in both pre-study and post-study phases. Additionally, sentiment scores regarding Mike, collected at two different points in the sentiment collection phase, were incorporated. The assigned label for this data pair was ``0'', indicative of positive opinion polarization.

\subsection{Experimental Setup}
%不同信号预测，加EEG效果更好
%detection/prediction   之前Nature那篇看看
\subsubsection{Feature Extraction}
\label{DE}
During the pre-study and post-study, continuous collection of Electroencephalogram (EEG) physiological signals was conducted on each participant. In this study, we utilized a 64-channel Quik-Cap (Compumedics NeuroScan) for EEG signal acquisition, with electrode placement following the International 10-20 system~\cite{homan1987cerebral}. The collected EEG data underwent preprocessing for further analysis, which included re-referencing to the average mastoid, baseline correction, low-pass filtering at 50 Hz, high-pass filtering at 0.5 Hz, and artifact removal~\cite{huizenga2002spatiotemporal}.

To utilize EEG data as features for training our model, we opted for Differential Entropy (DE) characteristics. Differential entropy (DE) is a crucial tool in assessing EEG signals~\cite{duan2013differential}. The computation of DE began with estimating the power spectral density (PSD), denoted as $P(f)$, where $f$ represents frequency. This estimation was conducted using Welch's method~\cite{welch1967use}, which employs a moving window technique. The length of this window was set to twice the inverse of the frequency band's lower limit, with the system operating at a sampling rate of 1000 Hz. We calculated DE as follows:

\begin{equation}
    DE = - \int P(f) \log(P(f)) \, df
\end{equation}

\subsubsection{Feature selection}
We have implemented three input user signals and their combinations as distinct input features:
(1) Behavior response Collected in our platform, including ``liking'' and ``viewing duration ratio''.
(2) EEG features in terms of Differential Entropy (DE) collected at every
electrode across five passbands, following the method described
in Section~\ref{DE}.
(3) Explicit response in terms of sentiment scores, on a five-point scale (1-5), collected from two questionnaires.
In our predictive modeling, feature selection was essential due to the substantially larger dimensions of EEG features compared to others. For each user-personage pair, we included four EEG features. 
These features consist of EEG signals collected during four specific conditions: viewing a positive/negative video~(denoted as pos-videos and neg-videos) in the pre-study/post-study.
The dimensionality of each EEG feature was 30 * 62 * 5 (i.e., \#time windows * \#electrodes * \#bands). To leverage the most effective EEG features, we selected the alpha frequency band, which is closely associated with emotions and cognitive functions~\cite{klimesch1999eeg}. 
We selected the "FPz" feature in the alpha band guided by Dixon et al. (2017)~\cite{dixon2017emotion}, who emphasized its relevance to emotional processing and focus. 
Anticipating the design simplicity required for large-scale commercial devices, our work demonstrates that single-channel predictions are on par with multi-channel ones, hence our emphasis on single-channel outcomes.
To represent user-personage level information, we averaged each feature over time (i.e., average (30,62,5) to (62,5)), followed by the selection of specific frequency bands and electrodes.

\subsubsection{Classification Model}
In our research, we explored various classical models before ultimately selecting the XGBoost model~\cite{chen2016xgboost}, recognized for its unique characteristics and benefits.
We used the default parameter settings of XGBoost based on the Python scikit-learn library.
XGBoost, an acronym for eXtreme Gradient Boosting, represents an advanced application of gradient boosting algorithms. 
This method's flexibility is evident in its applicability to both regression and classification tasks.

Our decision to employ XGBoost was influenced by its consistently strong performance and stability across diverse datasets. Notably, XGBoost excels in efficiency and effectiveness, aligning seamlessly with the specific requirements of our study. The high cost and time constraints associated with EEG data collection resulted in a relatively limited set of features. Consequently, we refrained from employing more complex models. Future research could potentially expand the dataset and explore the application of more sophisticated models.

\subsubsection{Data Splitting Protocol}
In Section~\ref{5.1}, we clearly defined the relationship between each feature type and user-personage pair and used these features to construct our dataset. 
This process yielded data for 230 user-personage pairs. We then aggregated these datasets and applied a five-fold cross-validation method to randomly split the data into training and test sets. Averaging the results from the test sets enhanced the reliability of our findings.

\begin{table}[t]
\centering
\caption{Performance of prediction by using different input features and their combinations.}
\label{tab:pred_results}
\small
\begin{tabular}{lccc}
\toprule
\multicolumn{1}{c}{\textbf{Input Features}} & \multicolumn{1}{c}{\textbf{ACC}} & \multicolumn{1}{c}{\textbf{F1}} & \multicolumn{1}{c}{\textbf{AUC}} \\
\midrule
Like & $0.557 \pm 0.015$ & $0.485 \pm 0.021$ & $0.548 \pm 0.023$ \\
%Duration & $0.463 \pm 0.017$ & $0.461 \pm 0.016$ & $0.432 \pm 0.026$ \\
Like+Duration & $0.508 \pm 0.021$ & $0.494 \pm 0.025$ & $0.552 \pm 0.020$ \\
Sentiment & $0.592 \pm 0.029$ & $0.588 \pm 0.028$ & $0.662 \pm 0.038$ \\
EEG & $0.641 \pm 0.024$ & $0.639 \pm 0.024$ & $0.672 \pm 0.029$ \\
Sentiment+EEG & $0.622 \pm 0.019$ & $0.620 \pm 0.019$ & $0.677 \pm 0.016$ \\
Like+EEG & $0.627 \pm 0.022$ & $0.623 \pm 0.023$ & $0.670 \pm 0.030$ \\
Like+Sentiment & $0.607 \pm 0.023$ & $0.598 \pm 0.023$ & $0.630 \pm 0.034$ \\
Like+Sentiment+EEG & $\mathbf{0.647 \pm 0.019}$ & $\mathbf{0.644 \pm 0.020}$ & $\mathbf{0.687 \pm 0.021}$ \\
\bottomrule
\end{tabular}
\end{table}

%\vspace{-5mm}

\subsection{Results}
%From Table~\ref{tab:pred_results}, we have the following observations:
The classification results are detailed in Table~\ref{tab:pred_results}, evaluated using metrics such as the Area Under the Curve (AUC), F1 Score (F1), and Accuracy (ACC)~\cite{huang2005using}. 

First, when relying solely on individual behavioral metrics (i.e., like or viewing duration), the prediction performance was notably weak, with AUCs only reaching 0.548 (like) and 0.552 (like+duration). 
Regardless, these AUC performances are marginally better than the 0.5 baseline which represents random chance, indicating that individual behavioral metrics have limited effectiveness in predicting the exposure to OP.

Second, \textbf{using EEG signals as a standalone feature yielded superior predictive results compared to traditional behavioral features.}
The performance metrics indicate considerable improvements over behavioral-based predictions, demonstrating EEG's effectiveness in predicting users' exposure to different types of polarized short videos. Furthermore, when sentiment annotations were independently used as features, they showed enhanced performance compared to behavior-based models. As a form of explicit feedback, the AUC performance of sentiment annotations is close to that of EEG, suggesting their potential utility in enhancing predictive accuracy.
%but its ACC and F1 scores are not as high as those of EEG.

Further, combining EEG and like features (we use like instead of the combination of the like and duration for better performance), surpassing the results obtained using only behavioral (i.e., like) features in ACC (from 0.557 to 0.627) and AUC (from 0.548 to 0.670). 
Combining EEG and sentiment features also showed better performance than using sentiment alone.
%with an ACC of 0.622, an F1 score of 0.657, and an AUC of 0.677. 
Moreover, the combination of like and sentiment produced better performance in ACC than using like or sentiment alone.

Last, \textbf{a combination of like, sentiment, and EEG features led to the best performance, with an ACC of 0.647, an F1 score of 0.644, and an AUC of 0.687}. The high performance reflects a significant likelihood of accurately predicting the exposure to OP. 
Combined results demonstrate that EEG features substantially contribute to predicting exposure to OP.

\paragraph{Answer to \textbf{RQ3}:} The experimental findings in the polarity classification task indicate that EEG signals, along with explicit responses (i.e., sentiment annotations), can be effectively utilized to classify polarity. 
This method demonstrates superior performance compared to traditional behavioral information. 
While we usually believe that explicit user signals are effective, this study reveals that EEG could be even more effective as explicit signals are suffer from the fact that users are sometimes unaware of OP.
Additionally, the integration of EEG and explicit responses with behavioral data can further enhance the performance of polarity classification.

\section{related work}
\subsection{Opinion Polarization}
Opinion Polarization refers to the increasing divergence in viewpoints among individuals or groups in society on specific issues, resulting in distinct clusters of opinions. This concept, explored in depth by \citet{mccoy2018polarization}, is often discussed in the context of "Us" versus "Them," categorizing people into groups based on their opinions. This polarization includes not just varying opinions, but also how individuals align with opposing opinion groups, as explained by Koudenburg and Kashima, 2021~\cite{koudenburg2021new}, and Turner and Hogg, 1987~\cite{hogg1987intergroup}. Theoretical and empirical studies, like those by Esteban and Ray, 1994~\cite{esteban1994measurement}, and Duclos et al., 2004~\cite{duclos2006polarization}, have focused on the psychological distances within opinion distributions and their impact on societal dynamics and conflicts.

%\subsection{Short Video Recommendation}
In the short video scenario, X Cheng et al., 2007~\cite{cheng2007understanding} found that videos have strong correlations with each other on YouTube.
YH Wang et al.,2017~\cite{wang2019causes} analyzed the characteristics and causes of the internet community of short video platforms.
X Cheng et al., 2009~\cite{cheng2009nettube} explored the clustering in social networks for short video sharing.
Y Gao et al., 2023~\cite{gao2023echo} found that the gathering of users into homogeneous groups dominates online interactions on Douyin and Bilibili (two leading short video platforms in China).
Xinyue Cao et al., 2021~\cite{cao2021destination} found that the narrative of short videos could promote brand attitude.
However, there is a lack of research on the effects of OP environment, as measured by EEG signals. 

\subsection{User Signals in IR}
User signals play an important role in enhancing the effectiveness and accuracy of information systems via methods such as user intent modeling~\cite{yang2023debiased}, relevance feedback~\cite{ye2023relevance,63white2002use}, and click models~\cite{chuklin2022click}.
User signals in \ac{IR} systems can be broadly categorized into two groups, explicit signals and implicit signals.
Implicit signals are indirect indications of user preferences or behavior, inferred from their behaviors such as clicks~\cite{63white2002use}, dwell time~\cite{48morita1994information}, eye-tracking~\cite{2akuma2022eye}, etc.
Explicit signals are direct and clear indications of user preferences or intentions, such as search queries entered into a search engine, likes or ratings to recommended items, etc.

Among the user signals in IR, \ac{EEG} signals have recently demonstrated remarkable capabilities in understanding user interactions in IR systems across various settings~\cite{ye2024brain}.
For example, detecting cognitive activities~\cite{2020decade, minas2014putting}, understanding information need~\cite{moshfeghi2019towards,moshfeghi2016understanding}, and predicting relevance judgment~\cite{ye2023relevance,ye2022don}.
Differently from traditional implicit signals and explicit signals, \ac{EEG} signals are not easily categorized as either explicit or implicit signals, because they are collected implicitly but directly reflect the user's mental activity.
Existing research has shown its superiority over traditional implicit signals in terms of accuracy~\cite{ye2022don}. 

This paper explores the effectiveness of various user signals in the context of OP.
Conventionally, behavior signals, especially clicks as implicit signals and likes as explicit signals are collected to measure and quantify OP.
However, these signals, as merely external manifestations of OP, might fall short of providing evidence of the impacts of OP on a user's cognitive perception of items on social platforms.
This paper further explores OP utilizing explicit responses related to sentiment, along with feedback based on EEG responses.

\section{Discussions and conclusions}
Understanding the effect of opinion polarization in short video browsing is crucial for shaping public opinion~\cite{yarchi2021political} and promoting healthier online environments~\cite{druckman2021affective}. 
To our knowledge, our study is the first to explore the impact of OP in short video browsing using EEG and sentiment annotations. 
We conducted a three-stage user study to gather explicit and implicit feedback during short video browsing. 
The feedback included EEG data, sentiment annotations, and behavioral data (likes and duration). 
We analyzed the collected data to address our research questions (RQs).
We found that OP exposure significantly affected user sentiments toward specific personage in short video browsing scenarios. If people encounter polarised opinions on social media for a prolonged time, it could influence their sentiments and, hence, broader societal perspectives on public figures and political issues.
Also, we found correlations between EEG signals and the direction of OP. 
This suggests that EEG can detect the type of OP exposure, revealing opinion polarization even without the users' awareness.
% Our experiments show that EEG and sentiment features can be used to predict OP exposure, and their performance outperformed behavioral features. 
% Despite the small dataset and simple method, the results are promising. 
% This indicates our paradigm for collecting EEG, sentiment, and behavioral data to detect OP is effective.

While the goal of this study is to detect and quantify the impact of exposure to polarised content on social media users on an individual level to shed more light on OP's social impact. 
However, our methods can also be used in other situations where people are exposed to informational content, such as evaluating how information content impacts students' sentiments in education scenarios. 
Our study reveals that brain signals act as a more accurate and direct measurement of users' implicit responses to OP, revealing impacts that traditional behavioral data might be unable to capture. 
With advancements in lightweight brain data collection, brain-related devices can be applied in IR, virtual reality, disabled services, and NLP applications like intelligent assistants, offering enhanced information quality and real-time feedback in these contexts. 
Based on specific brain signals, those new technologies may provide the users with more relevant, diverse, or balanced content, depending on the usage scenario. 
We hope that our study could inspire researchers from areas of information science, social sciences, and psychology to use EEG to understand knowledge acquisition and opinion formation processes.

Limitations of our study may lead to future directions for research:
(1) Our study's participant scale was limited. We recruited 24 participants due to the high costs of EEG data collection.
Collecting larger-scale EEG and sentiment data could potentially improve the performance of the OP prediction models.
(2) There's a gap between our provided short video browsing platforms and those used by users every day. 
It would be beneficial, if possible, to conduct field studies with EEG devices on real-world short video recommendation systems.
(3) The OP prediction model we built is relatively simple. Exploring more complex methods such as generative AI models might be a fruitful direction.
\end{CJK}
\clearpage
\bibliographystyle{ACM-Reference-Format}
\normalem
% \balance
\bibliography{references}


\begin{thebibliography}{60}

%%% the Web address.

\ifx \showCODEN    \undefined \def \showCODEN     #1{\unskip}     \fi
\ifx \showDOI      \undefined \def \showDOI       #1{#1}\fi
\ifx \showISBNx    \undefined \def \showISBNx     #1{\unskip}     \fi
\ifx \showISBNxiii \undefined \def \showISBNxiii  #1{\unskip}     \fi
\ifx \showISSN     \undefined \def \showISSN      #1{\unskip}     \fi
\ifx \showLCCN     \undefined \def \showLCCN      #1{\unskip}     \fi
\ifx \shownote     \undefined \def \shownote      #1{#1}          \fi
\ifx \showarticletitle \undefined \def \showarticletitle #1{#1}   \fi
\ifx \showURL      \undefined \def \showURL       {\relax}        \fi
% invisible to TeX
\providecommand\bibfield[2]{#2}
\providecommand\bibinfo[2]{#2}
\providecommand\natexlab[1]{#1}
\providecommand\showeprint[2][]{arXiv:#2}

\bibitem[\protect\citeauthoryear{Abebe, Kleinberg, Parkes, and Tsourakakis}{Abebe et~al\mbox{.}}{2018}]%
        {abebe2018opinion}
\bibfield{author}{\bibinfo{person}{Rediet Abebe}, \bibinfo{person}{Jon Kleinberg}, \bibinfo{person}{David Parkes}, {and} \bibinfo{person}{Charalampos~E Tsourakakis}.} \bibinfo{year}{2018}\natexlab{}.
\newblock \showarticletitle{Opinion dynamics with varying susceptibility to persuasion}. In \bibinfo{booktitle}{\emph{Proceedings of the 24th ACM SIGKDD International Conference on Knowledge Discovery \& Data Mining}}. \bibinfo{pages}{1089--1098}.
\newblock


\bibitem[\protect\citeauthoryear{Akuma}{Akuma}{2022}]%
        {2akuma2022eye}
\bibfield{author}{\bibinfo{person}{Stephen Akuma}.} \bibinfo{year}{2022}\natexlab{}.
\newblock \showarticletitle{Eye Gaze Relevance Feedback Indicators for Information Retrieval}.
\newblock \bibinfo{journal}{\emph{International Journal of Intelligent Systems and Applications (IJISA)}} \bibinfo{volume}{14}, \bibinfo{number}{1} (\bibinfo{year}{2022}), \bibinfo{pages}{57--65}.
\newblock


\bibitem[\protect\citeauthoryear{Asker and Dinas}{Asker and Dinas}{2019}]%
        {asker2019thinking}
\bibfield{author}{\bibinfo{person}{David Asker} {and} \bibinfo{person}{Elias Dinas}.} \bibinfo{year}{2019}\natexlab{}.
\newblock \showarticletitle{Thinking fast and furious: Emotional intensity and opinion polarization in online media}.
\newblock \bibinfo{journal}{\emph{Public Opinion Quarterly}} \bibinfo{volume}{83}, \bibinfo{number}{3} (\bibinfo{year}{2019}), \bibinfo{pages}{487--509}.
\newblock


\bibitem[\protect\citeauthoryear{Cao, Qu, Liu, and Hu}{Cao et~al\mbox{.}}{2021}]%
        {cao2021destination}
\bibfield{author}{\bibinfo{person}{Xinyue Cao}, \bibinfo{person}{Zhirui Qu}, \bibinfo{person}{Yan Liu}, {and} \bibinfo{person}{JiaJing Hu}.} \bibinfo{year}{2021}\natexlab{}.
\newblock \showarticletitle{How the destination short video affects the customers' attitude: The role of narrative transportation}.
\newblock \bibinfo{journal}{\emph{Journal of Retailing and Consumer Services}}  \bibinfo{volume}{62} (\bibinfo{year}{2021}), \bibinfo{pages}{102672}.
\newblock


\bibitem[\protect\citeauthoryear{Chen}{Chen}{2021}]%
        {chen2021exploration}
\bibfield{author}{\bibinfo{person}{Minmin Chen}.} \bibinfo{year}{2021}\natexlab{}.
\newblock \showarticletitle{Exploration in recommender systems}. In \bibinfo{booktitle}{\emph{Proceedings of the 15th ACM Conference on Recommender Systems}}. \bibinfo{pages}{551--553}.
\newblock


\bibitem[\protect\citeauthoryear{Chen, Wang, Xu, Le, Sharma, Richardson, Wu, and Chi}{Chen et~al\mbox{.}}{2021}]%
        {chen2021values}
\bibfield{author}{\bibinfo{person}{Minmin Chen}, \bibinfo{person}{Yuyan Wang}, \bibinfo{person}{Can Xu}, \bibinfo{person}{Ya Le}, \bibinfo{person}{Mohit Sharma}, \bibinfo{person}{Lee Richardson}, \bibinfo{person}{Su-Lin Wu}, {and} \bibinfo{person}{Ed Chi}.} \bibinfo{year}{2021}\natexlab{}.
\newblock \showarticletitle{Values of user exploration in recommender systems}. In \bibinfo{booktitle}{\emph{Proceedings of the 15th ACM Conference on Recommender Systems}}. \bibinfo{pages}{85--95}.
\newblock


\bibitem[\protect\citeauthoryear{Chen, Qiu, Zhao, Han, He, Siponen, Mou, and Xiao}{Chen et~al\mbox{.}}{2022}]%
        {chen2022more}
\bibfield{author}{\bibinfo{person}{Sihua Chen}, \bibinfo{person}{Han Qiu}, \bibinfo{person}{Shifei Zhao}, \bibinfo{person}{Yuyu Han}, \bibinfo{person}{Wei He}, \bibinfo{person}{Mikko Siponen}, \bibinfo{person}{Jian Mou}, {and} \bibinfo{person}{Hua Xiao}.} \bibinfo{year}{2022}\natexlab{}.
\newblock \showarticletitle{When more is less: The other side of artificial intelligence recommendation}.
\newblock \bibinfo{journal}{\emph{Journal of Management Science and Engineering}} \bibinfo{volume}{7}, \bibinfo{number}{2} (\bibinfo{year}{2022}), \bibinfo{pages}{213--232}.
\newblock


\bibitem[\protect\citeauthoryear{Chen and Guestrin}{Chen and Guestrin}{2016}]%
        {chen2016xgboost}
\bibfield{author}{\bibinfo{person}{Tianqi Chen} {and} \bibinfo{person}{Carlos Guestrin}.} \bibinfo{year}{2016}\natexlab{}.
\newblock \showarticletitle{Xgboost: A scalable tree boosting system}. In \bibinfo{booktitle}{\emph{Proceedings of the 22nd acm sigkdd international conference on knowledge discovery and data mining}}. \bibinfo{pages}{785--794}.
\newblock


\bibitem[\protect\citeauthoryear{Cheng, Dale, and Liu}{Cheng et~al\mbox{.}}{2007}]%
        {cheng2007understanding}
\bibfield{author}{\bibinfo{person}{Xu Cheng}, \bibinfo{person}{Cameron Dale}, {and} \bibinfo{person}{Jiangchuan Liu}.} \bibinfo{year}{2007}\natexlab{}.
\newblock \showarticletitle{Understanding the characteristics of internet short video sharing: YouTube as a case study}.
\newblock \bibinfo{journal}{\emph{arXiv preprint arXiv:0707.3670}} (\bibinfo{year}{2007}).
\newblock


\bibitem[\protect\citeauthoryear{Cheng and Liu}{Cheng and Liu}{2009}]%
        {cheng2009nettube}
\bibfield{author}{\bibinfo{person}{Xu Cheng} {and} \bibinfo{person}{Jiangchuan Liu}.} \bibinfo{year}{2009}\natexlab{}.
\newblock \showarticletitle{Nettube: Exploring social networks for peer-to-peer short video sharing}. In \bibinfo{booktitle}{\emph{IEEE INFOCOM 2009}}. IEEE, \bibinfo{pages}{1152--1160}.
\newblock


\bibitem[\protect\citeauthoryear{Chitra and Musco}{Chitra and Musco}{2020}]%
        {chitra2020analyzing}
\bibfield{author}{\bibinfo{person}{Uthsav Chitra} {and} \bibinfo{person}{Christopher Musco}.} \bibinfo{year}{2020}\natexlab{}.
\newblock \showarticletitle{Analyzing the impact of filter bubbles on social network polarization}. In \bibinfo{booktitle}{\emph{Proceedings of the 13th International Conference on Web Search and Data Mining}}. \bibinfo{pages}{115--123}.
\newblock


\bibitem[\protect\citeauthoryear{Chuklin, Markov, and De~Rijke}{Chuklin et~al\mbox{.}}{2022}]%
        {chuklin2022click}
\bibfield{author}{\bibinfo{person}{Aleksandr Chuklin}, \bibinfo{person}{Ilya Markov}, {and} \bibinfo{person}{Maarten De~Rijke}.} \bibinfo{year}{2022}\natexlab{}.
\newblock \bibinfo{booktitle}{\emph{Click models for web search}}.
\newblock \bibinfo{publisher}{Springer Nature}.
\newblock


\bibitem[\protect\citeauthoryear{Criss, Michaels, Solomon, Allen, and Nguyen}{Criss et~al\mbox{.}}{2021}]%
        {criss2021twitter}
\bibfield{author}{\bibinfo{person}{Shaniece Criss}, \bibinfo{person}{Eli~K Michaels}, \bibinfo{person}{Kamra Solomon}, \bibinfo{person}{Amani~M Allen}, {and} \bibinfo{person}{Thu~T Nguyen}.} \bibinfo{year}{2021}\natexlab{}.
\newblock \showarticletitle{Twitter fingers and echo chambers: Exploring expressions and experiences of online racism using twitter}.
\newblock \bibinfo{journal}{\emph{Journal of Racial and Ethnic Health Disparities}}  \bibinfo{volume}{8} (\bibinfo{year}{2021}), \bibinfo{pages}{1322--1331}.
\newblock


\bibitem[\protect\citeauthoryear{Dixon, Thiruchselvam, Todd, and Christoff}{Dixon et~al\mbox{.}}{2017}]%
        {dixon2017emotion}
\bibfield{author}{\bibinfo{person}{Matthew~L Dixon}, \bibinfo{person}{Ravi Thiruchselvam}, \bibinfo{person}{Rebecca Todd}, {and} \bibinfo{person}{Kalina Christoff}.} \bibinfo{year}{2017}\natexlab{}.
\newblock \showarticletitle{Emotion and the prefrontal cortex: An integrative review.}
\newblock \bibinfo{journal}{\emph{Psychological bulletin}} \bibinfo{volume}{143}, \bibinfo{number}{10} (\bibinfo{year}{2017}), \bibinfo{pages}{1033}.
\newblock


\bibitem[\protect\citeauthoryear{Druckman, Klar, Krupnikov, Levendusky, and Ryan}{Druckman et~al\mbox{.}}{2021}]%
        {druckman2021affective}
\bibfield{author}{\bibinfo{person}{James~N Druckman}, \bibinfo{person}{Samara Klar}, \bibinfo{person}{Yanna Krupnikov}, \bibinfo{person}{Matthew Levendusky}, {and} \bibinfo{person}{John~Barry Ryan}.} \bibinfo{year}{2021}\natexlab{}.
\newblock \showarticletitle{Affective polarization, local contexts and public opinion in America}.
\newblock \bibinfo{journal}{\emph{Nature human behaviour}} \bibinfo{volume}{5}, \bibinfo{number}{1} (\bibinfo{year}{2021}), \bibinfo{pages}{28--38}.
\newblock


\bibitem[\protect\citeauthoryear{Duan, Zhu, and Lu}{Duan et~al\mbox{.}}{2013}]%
        {duan2013differential}
\bibfield{author}{\bibinfo{person}{Ruo-Nan Duan}, \bibinfo{person}{Jia-Yi Zhu}, {and} \bibinfo{person}{Bao-Liang Lu}.} \bibinfo{year}{2013}\natexlab{}.
\newblock \showarticletitle{Differential entropy feature for EEG-based emotion classification}. In \bibinfo{booktitle}{\emph{2013 6th International IEEE/EMBS Conference on Neural Engineering (NER)}}. IEEE, \bibinfo{pages}{81--84}.
\newblock


\bibitem[\protect\citeauthoryear{Duclos, Esteban, and Ray}{Duclos et~al\mbox{.}}{2006}]%
        {duclos2006polarization}
\bibfield{author}{\bibinfo{person}{Jean-Yves Duclos}, \bibinfo{person}{Joan Esteban}, {and} \bibinfo{person}{Debraj Ray}.} \bibinfo{year}{2006}\natexlab{}.
\newblock \showarticletitle{Polarization: concepts, measurement, estimation}.
\newblock In \bibinfo{booktitle}{\emph{The Social Economics of Poverty}}. \bibinfo{publisher}{Routledge}, \bibinfo{pages}{54--102}.
\newblock


\bibitem[\protect\citeauthoryear{Einav, Allen, Gur, Maaravi, and Ravner}{Einav et~al\mbox{.}}{2022}]%
        {einav2022bursting}
\bibfield{author}{\bibinfo{person}{Gali Einav}, \bibinfo{person}{Ofir Allen}, \bibinfo{person}{Tamar Gur}, \bibinfo{person}{Yossi Maaravi}, {and} \bibinfo{person}{Daniel Ravner}.} \bibinfo{year}{2022}\natexlab{}.
\newblock \showarticletitle{Bursting filter bubbles in a digital age: Opening minds and reducing opinion polarization through digital platforms}.
\newblock \bibinfo{journal}{\emph{Technology in Society}}  \bibinfo{volume}{71} (\bibinfo{year}{2022}), \bibinfo{pages}{102136}.
\newblock


\bibitem[\protect\citeauthoryear{Ertl, Hildebrandt, Ourina, Leicht, and Mulert}{Ertl et~al\mbox{.}}{2013}]%
        {ertl2013emotion}
\bibfield{author}{\bibinfo{person}{Matthias Ertl}, \bibinfo{person}{Maria Hildebrandt}, \bibinfo{person}{Kristina Ourina}, \bibinfo{person}{Gregor Leicht}, {and} \bibinfo{person}{Christoph Mulert}.} \bibinfo{year}{2013}\natexlab{}.
\newblock \showarticletitle{Emotion regulation by cognitive reappraisal—the role of frontal theta oscillations}.
\newblock \bibinfo{journal}{\emph{NeuroImage}}  \bibinfo{volume}{81} (\bibinfo{year}{2013}), \bibinfo{pages}{412--421}.
\newblock


\bibitem[\protect\citeauthoryear{Esteban and Ray}{Esteban and Ray}{1994}]%
        {esteban1994measurement}
\bibfield{author}{\bibinfo{person}{Joan-Maria Esteban} {and} \bibinfo{person}{Debraj Ray}.} \bibinfo{year}{1994}\natexlab{}.
\newblock \showarticletitle{On the measurement of polarization}.
\newblock \bibinfo{journal}{\emph{Econometrica: Journal of the Econometric Society}} (\bibinfo{year}{1994}), \bibinfo{pages}{819--851}.
\newblock


\bibitem[\protect\citeauthoryear{Gao, Liu, and Gao}{Gao et~al\mbox{.}}{2023}]%
        {gao2023echo}
\bibfield{author}{\bibinfo{person}{Yichang Gao}, \bibinfo{person}{Fengming Liu}, {and} \bibinfo{person}{Lei Gao}.} \bibinfo{year}{2023}\natexlab{}.
\newblock \showarticletitle{Echo chamber effects on short video platforms}.
\newblock \bibinfo{journal}{\emph{Scientific Reports}} \bibinfo{volume}{13}, \bibinfo{number}{1} (\bibinfo{year}{2023}), \bibinfo{pages}{6282}.
\newblock


\bibitem[\protect\citeauthoryear{Groshek and Koc-Michalska}{Groshek and Koc-Michalska}{2017}]%
        {groshek2017helping}
\bibfield{author}{\bibinfo{person}{Jacob Groshek} {and} \bibinfo{person}{Karolina Koc-Michalska}.} \bibinfo{year}{2017}\natexlab{}.
\newblock \showarticletitle{Helping populism win? Social media use, filter bubbles, and support for populist presidential candidates in the 2016 US election campaign}.
\newblock \bibinfo{journal}{\emph{Information, Communication \& Society}} \bibinfo{volume}{20}, \bibinfo{number}{9} (\bibinfo{year}{2017}), \bibinfo{pages}{1389--1407}.
\newblock


\bibitem[\protect\citeauthoryear{He, Zhang, Sun, Li, Xie, Zhang, and Liu}{He et~al\mbox{.}}{2023}]%
        {he2023understanding}
\bibfield{author}{\bibinfo{person}{Zhiyu He}, \bibinfo{person}{Shaorun Zhang}, \bibinfo{person}{Peijie Sun}, \bibinfo{person}{Jiayu Li}, \bibinfo{person}{Xiaohui Xie}, \bibinfo{person}{Min Zhang}, {and} \bibinfo{person}{Yiqun Liu}.} \bibinfo{year}{2023}\natexlab{}.
\newblock \showarticletitle{Understanding User Immersion in Online Short Video Interaction}. In \bibinfo{booktitle}{\emph{Proceedings of the 32nd ACM International Conference on Information and Knowledge Management}}. \bibinfo{pages}{731--740}.
\newblock


\bibitem[\protect\citeauthoryear{Headley and Weinberger}{Headley and Weinberger}{2011}]%
        {headley2011gamma}
\bibfield{author}{\bibinfo{person}{Drew~B Headley} {and} \bibinfo{person}{Norman~M Weinberger}.} \bibinfo{year}{2011}\natexlab{}.
\newblock \showarticletitle{Gamma-band activation predicts both associative memory and cortical plasticity}.
\newblock \bibinfo{journal}{\emph{Journal of Neuroscience}} \bibinfo{volume}{31}, \bibinfo{number}{36} (\bibinfo{year}{2011}), \bibinfo{pages}{12748--12758}.
\newblock


\bibitem[\protect\citeauthoryear{Hogg and Turner}{Hogg and Turner}{1987}]%
        {hogg1987intergroup}
\bibfield{author}{\bibinfo{person}{Michael~A Hogg} {and} \bibinfo{person}{John~C Turner}.} \bibinfo{year}{1987}\natexlab{}.
\newblock \showarticletitle{Intergroup behaviour, self-stereotyping and the salience of social categories}.
\newblock \bibinfo{journal}{\emph{British Journal of Social Psychology}} \bibinfo{volume}{26}, \bibinfo{number}{4} (\bibinfo{year}{1987}), \bibinfo{pages}{325--340}.
\newblock


\bibitem[\protect\citeauthoryear{Holone}{Holone}{2016}]%
        {holone2016filter}
\bibfield{author}{\bibinfo{person}{Harald Holone}.} \bibinfo{year}{2016}\natexlab{}.
\newblock \showarticletitle{The filter bubble and its effect on online personal health information}.
\newblock \bibinfo{journal}{\emph{Croatian medical journal}} \bibinfo{volume}{57}, \bibinfo{number}{3} (\bibinfo{year}{2016}), \bibinfo{pages}{298}.
\newblock


\bibitem[\protect\citeauthoryear{Homan, Herman, and Purdy}{Homan et~al\mbox{.}}{1987}]%
        {homan1987cerebral}
\bibfield{author}{\bibinfo{person}{Richard~W Homan}, \bibinfo{person}{John Herman}, {and} \bibinfo{person}{Phillip Purdy}.} \bibinfo{year}{1987}\natexlab{}.
\newblock \showarticletitle{Cerebral location of international 10--20 system electrode placement}.
\newblock \bibinfo{journal}{\emph{Electroencephalography and clinical neurophysiology}} \bibinfo{volume}{66}, \bibinfo{number}{4} (\bibinfo{year}{1987}), \bibinfo{pages}{376--382}.
\newblock


\bibitem[\protect\citeauthoryear{Hosanagar, Fleder, Lee, and Buja}{Hosanagar et~al\mbox{.}}{2014}]%
        {hosanagar2014will}
\bibfield{author}{\bibinfo{person}{Kartik Hosanagar}, \bibinfo{person}{Daniel Fleder}, \bibinfo{person}{Dokyun Lee}, {and} \bibinfo{person}{Andreas Buja}.} \bibinfo{year}{2014}\natexlab{}.
\newblock \showarticletitle{Will the global village fracture into tribes? Recommender systems and their effects on consumer fragmentation}.
\newblock \bibinfo{journal}{\emph{Management Science}} \bibinfo{volume}{60}, \bibinfo{number}{4} (\bibinfo{year}{2014}), \bibinfo{pages}{805--823}.
\newblock


\bibitem[\protect\citeauthoryear{Huang and Ling}{Huang and Ling}{2005}]%
        {huang2005using}
\bibfield{author}{\bibinfo{person}{Jin Huang} {and} \bibinfo{person}{Charles~X Ling}.} \bibinfo{year}{2005}\natexlab{}.
\newblock \showarticletitle{Using AUC and accuracy in evaluating learning algorithms}.
\newblock \bibinfo{journal}{\emph{IEEE Transactions on knowledge and Data Engineering}} \bibinfo{volume}{17}, \bibinfo{number}{3} (\bibinfo{year}{2005}), \bibinfo{pages}{299--310}.
\newblock


\bibitem[\protect\citeauthoryear{Huizenga, De~Munck, Waldorp, and Grasman}{Huizenga et~al\mbox{.}}{2002}]%
        {huizenga2002spatiotemporal}
\bibfield{author}{\bibinfo{person}{Hilde~M Huizenga}, \bibinfo{person}{Jan~C De~Munck}, \bibinfo{person}{Lourens~J Waldorp}, {and} \bibinfo{person}{Raoul~PPP Grasman}.} \bibinfo{year}{2002}\natexlab{}.
\newblock \showarticletitle{Spatiotemporal EEG/MEG source analysis based on a parametric noise covariance model}.
\newblock \bibinfo{journal}{\emph{IEEE Transactions on Biomedical Engineering}} \bibinfo{volume}{49}, \bibinfo{number}{6} (\bibinfo{year}{2002}), \bibinfo{pages}{533--539}.
\newblock


\bibitem[\protect\citeauthoryear{Klimesch}{Klimesch}{1999}]%
        {klimesch1999eeg}
\bibfield{author}{\bibinfo{person}{Wolfgang Klimesch}.} \bibinfo{year}{1999}\natexlab{}.
\newblock \showarticletitle{EEG alpha and theta oscillations reflect cognitive and memory performance: a review and analysis}.
\newblock \bibinfo{journal}{\emph{Brain research reviews}} \bibinfo{volume}{29}, \bibinfo{number}{2-3} (\bibinfo{year}{1999}), \bibinfo{pages}{169--195}.
\newblock


\bibitem[\protect\citeauthoryear{Koudenburg, Kiers, and Kashima}{Koudenburg et~al\mbox{.}}{2021}]%
        {koudenburg2021new}
\bibfield{author}{\bibinfo{person}{Namkje Koudenburg}, \bibinfo{person}{Henk~AL Kiers}, {and} \bibinfo{person}{Yoshihisa Kashima}.} \bibinfo{year}{2021}\natexlab{}.
\newblock \showarticletitle{A new opinion polarization index developed by integrating expert judgments}.
\newblock \bibinfo{journal}{\emph{Frontiers in psychology}}  \bibinfo{volume}{12} (\bibinfo{year}{2021}), \bibinfo{pages}{738258}.
\newblock


\bibitem[\protect\citeauthoryear{Kumar, Sharma, Ramakrishnan, and Adarsh}{Kumar et~al\mbox{.}}{2021}]%
        {kumar2021increased}
\bibfield{author}{\bibinfo{person}{G~Pradeep Kumar}, \bibinfo{person}{Kanishka Sharma}, \bibinfo{person}{AG Ramakrishnan}, {and} \bibinfo{person}{A Adarsh}.} \bibinfo{year}{2021}\natexlab{}.
\newblock \showarticletitle{Increased entropy of gamma oscillations in the frontal region during meditation}. In \bibinfo{booktitle}{\emph{2021 43rd Annual International Conference of the IEEE Engineering in Medicine \& Biology Society (EMBC)}}. IEEE, \bibinfo{pages}{787--790}.
\newblock


\bibitem[\protect\citeauthoryear{Landis and Koch}{Landis and Koch}{1977}]%
        {landis1977measurement}
\bibfield{author}{\bibinfo{person}{J~Richard Landis} {and} \bibinfo{person}{Gary~G Koch}.} \bibinfo{year}{1977}\natexlab{}.
\newblock \showarticletitle{The measurement of observer agreement for categorical data}.
\newblock \bibinfo{journal}{\emph{biometrics}} (\bibinfo{year}{1977}), \bibinfo{pages}{159--174}.
\newblock


\bibitem[\protect\citeauthoryear{Lee}{Lee}{2016}]%
        {lee2016impact}
\bibfield{author}{\bibinfo{person}{Francis~LF Lee}.} \bibinfo{year}{2016}\natexlab{}.
\newblock \showarticletitle{Impact of social media on opinion polarization in varying times}.
\newblock \bibinfo{journal}{\emph{Communication and the Public}} \bibinfo{volume}{1}, \bibinfo{number}{1} (\bibinfo{year}{2016}), \bibinfo{pages}{56--71}.
\newblock


\bibitem[\protect\citeauthoryear{Magnus and Van~der Holst}{Magnus and Van~der Holst}{1987}]%
        {magnus1987zeta}
\bibfield{author}{\bibinfo{person}{Otto Magnus} {and} \bibinfo{person}{Marian Van~der Holst}.} \bibinfo{year}{1987}\natexlab{}.
\newblock \showarticletitle{Zeta waves: a special type of slow delta waves}.
\newblock \bibinfo{journal}{\emph{Electroencephalography and clinical neurophysiology}} \bibinfo{volume}{67}, \bibinfo{number}{2} (\bibinfo{year}{1987}), \bibinfo{pages}{140--146}.
\newblock


\bibitem[\protect\citeauthoryear{Martinovic}{Martinovic}{2018}]%
        {martinovic2018exploring}
\bibfield{author}{\bibinfo{person}{Mario Martinovic}.} \bibinfo{year}{2018}\natexlab{}.
\newblock \emph{\bibinfo{title}{Exploring the Effect of Search Engine Personalization on Politically Biased Search Results}}.
\newblock {B.S.} thesis. \bibinfo{school}{University of Twente}.
\newblock


\bibitem[\protect\citeauthoryear{M{\"a}s and Flache}{M{\"a}s and Flache}{2013}]%
        {mas2013differentiation}
\bibfield{author}{\bibinfo{person}{Michael M{\"a}s} {and} \bibinfo{person}{Andreas Flache}.} \bibinfo{year}{2013}\natexlab{}.
\newblock \showarticletitle{Differentiation without distancing. Explaining bi-polarization of opinions without negative influence}.
\newblock \bibinfo{journal}{\emph{PloS one}} \bibinfo{volume}{8}, \bibinfo{number}{11} (\bibinfo{year}{2013}), \bibinfo{pages}{e74516}.
\newblock


\bibitem[\protect\citeauthoryear{McCoy, Rahman, and Somer}{McCoy et~al\mbox{.}}{2018}]%
        {mccoy2018polarization}
\bibfield{author}{\bibinfo{person}{Jennifer McCoy}, \bibinfo{person}{Tahmina Rahman}, {and} \bibinfo{person}{Murat Somer}.} \bibinfo{year}{2018}\natexlab{}.
\newblock \showarticletitle{Polarization and the global crisis of democracy: Common patterns, dynamics, and pernicious consequences for democratic polities}.
\newblock \bibinfo{journal}{\emph{American Behavioral Scientist}} \bibinfo{volume}{62}, \bibinfo{number}{1} (\bibinfo{year}{2018}), \bibinfo{pages}{16--42}.
\newblock


\bibitem[\protect\citeauthoryear{McCright and Dunlap}{McCright and Dunlap}{2011}]%
        {mccright2011politicization}
\bibfield{author}{\bibinfo{person}{Aaron~M McCright} {and} \bibinfo{person}{Riley~E Dunlap}.} \bibinfo{year}{2011}\natexlab{}.
\newblock \showarticletitle{The politicization of climate change and polarization in the American public's views of global warming, 2001--2010}.
\newblock \bibinfo{journal}{\emph{The Sociological Quarterly}} \bibinfo{volume}{52}, \bibinfo{number}{2} (\bibinfo{year}{2011}), \bibinfo{pages}{155--194}.
\newblock


\bibitem[\protect\citeauthoryear{Merrill~Jr and Rubenking}{Merrill~Jr and Rubenking}{2019}]%
        {merrill2019go}
\bibfield{author}{\bibinfo{person}{Kelly Merrill~Jr} {and} \bibinfo{person}{Bridget Rubenking}.} \bibinfo{year}{2019}\natexlab{}.
\newblock \showarticletitle{Go long or go often: Influences on binge watching frequency and duration among college students}.
\newblock \bibinfo{journal}{\emph{Social Sciences}} \bibinfo{volume}{8}, \bibinfo{number}{1} (\bibinfo{year}{2019}), \bibinfo{pages}{10}.
\newblock


\bibitem[\protect\citeauthoryear{Minas, Potter, Dennis, Bartelt, and Bae}{Minas et~al\mbox{.}}{2014}]%
        {minas2014putting}
\bibfield{author}{\bibinfo{person}{Randall~K Minas}, \bibinfo{person}{Robert~F Potter}, \bibinfo{person}{Alan~R Dennis}, \bibinfo{person}{Valerie Bartelt}, {and} \bibinfo{person}{Soyoung Bae}.} \bibinfo{year}{2014}\natexlab{}.
\newblock \showarticletitle{Putting on the thinking cap: using NeuroIS to understand information processing biases in virtual teams}.
\newblock \bibinfo{journal}{\emph{Journal of Management Information Systems}} \bibinfo{volume}{30}, \bibinfo{number}{4} (\bibinfo{year}{2014}), \bibinfo{pages}{49--82}.
\newblock


\bibitem[\protect\citeauthoryear{Morita and Shinoda}{Morita and Shinoda}{1994}]%
        {48morita1994information}
\bibfield{author}{\bibinfo{person}{Masahiro Morita} {and} \bibinfo{person}{Yoichi Shinoda}.} \bibinfo{year}{1994}\natexlab{}.
\newblock \showarticletitle{Information filtering based on user behavior analysis and best match text retrieval}. In \bibinfo{booktitle}{\emph{SIGIR’94: Proceedings of the Seventeenth Annual International ACM-SIGIR Conference on Research and Development in Information Retrieval, organised by Dublin City University}}. Springer, \bibinfo{pages}{272--281}.
\newblock


\bibitem[\protect\citeauthoryear{Moshfeghi, Pinto, Pollick, and Jose}{Moshfeghi et~al\mbox{.}}{2013}]%
        {moshfeghi2013understanding}
\bibfield{author}{\bibinfo{person}{Yashar Moshfeghi}, \bibinfo{person}{Luisa~R Pinto}, \bibinfo{person}{Frank~E Pollick}, {and} \bibinfo{person}{Joemon~M Jose}.} \bibinfo{year}{2013}\natexlab{}.
\newblock \showarticletitle{Understanding relevance: An fMRI study}. In \bibinfo{booktitle}{\emph{European conference on information retrieval}}. Springer, \bibinfo{pages}{14--25}.
\newblock


\bibitem[\protect\citeauthoryear{Moshfeghi, Triantafillou, and Pollick}{Moshfeghi et~al\mbox{.}}{2019}]%
        {moshfeghi2019towards}
\bibfield{author}{\bibinfo{person}{Yashar Moshfeghi}, \bibinfo{person}{Peter Triantafillou}, {and} \bibinfo{person}{Frank Pollick}.} \bibinfo{year}{2019}\natexlab{}.
\newblock \showarticletitle{Towards predicting a realisation of an information need based on brain signals}. In \bibinfo{booktitle}{\emph{The World Wide Web Conference}}. \bibinfo{pages}{1300--1309}.
\newblock


\bibitem[\protect\citeauthoryear{Moshfeghi, Triantafillou, and Pollick}{Moshfeghi et~al\mbox{.}}{2016}]%
        {moshfeghi2016understanding}
\bibfield{author}{\bibinfo{person}{Yashar Moshfeghi}, \bibinfo{person}{Peter Triantafillou}, {and} \bibinfo{person}{Frank~E Pollick}.} \bibinfo{year}{2016}\natexlab{}.
\newblock \showarticletitle{Understanding information need: An fMRI study}. In \bibinfo{booktitle}{\emph{Proceedings of the 39th International ACM SIGIR conference on Research and Development in Information Retrieval}}. \bibinfo{pages}{335--344}.
\newblock


\bibitem[\protect\citeauthoryear{Pariser}{Pariser}{2011}]%
        {pariser2011filter}
\bibfield{author}{\bibinfo{person}{Eli Pariser}.} \bibinfo{year}{2011}\natexlab{}.
\newblock \bibinfo{booktitle}{\emph{The filter bubble: What the Internet is hiding from you}}.
\newblock \bibinfo{publisher}{penguin UK}.
\newblock


\bibitem[\protect\citeauthoryear{Pinkosova, McGeown, and Moshfeghi}{Pinkosova et~al\mbox{.}}{2023}]%
        {pinkosova2023moderating}
\bibfield{author}{\bibinfo{person}{Zuzana Pinkosova}, \bibinfo{person}{William~J McGeown}, {and} \bibinfo{person}{Yashar Moshfeghi}.} \bibinfo{year}{2023}\natexlab{}.
\newblock \showarticletitle{Moderating effects of self-perceived knowledge in a relevance assessment task: An EEG study}.
\newblock \bibinfo{journal}{\emph{Computers in Human Behavior Reports}}  \bibinfo{volume}{11} (\bibinfo{year}{2023}), \bibinfo{pages}{100295}.
\newblock


\bibitem[\protect\citeauthoryear{Riedl, Fischer, L\'{e}ger, and Davis}{Riedl et~al\mbox{.}}{2020}]%
        {2020decade}
\bibfield{author}{\bibinfo{person}{Ren\'{e} Riedl}, \bibinfo{person}{Thomas Fischer}, \bibinfo{person}{Pierre-Majorique L\'{e}ger}, {and} \bibinfo{person}{Fred~D. Davis}.} \bibinfo{year}{2020}\natexlab{}.
\newblock \showarticletitle{A Decade of NeuroIS Research: Progress, Challenges, and Future Directions}.
\newblock \bibinfo{journal}{\emph{SIGMIS Database}} \bibinfo{volume}{51}, \bibinfo{number}{3} (\bibinfo{date}{jul} \bibinfo{year}{2020}), \bibinfo{pages}{13–54}.
\newblock
\showISSN{0095-0033}
\urldef\tempurl%
\url{https://doi.org/10.1145/3410977.3410980}
\showDOI{\tempurl}


\bibitem[\protect\citeauthoryear{Schober, Boer, and Schwarte}{Schober et~al\mbox{.}}{2018}]%
        {schober2018correlation}
\bibfield{author}{\bibinfo{person}{Patrick Schober}, \bibinfo{person}{Christa Boer}, {and} \bibinfo{person}{Lothar~A Schwarte}.} \bibinfo{year}{2018}\natexlab{}.
\newblock \showarticletitle{Correlation coefficients: appropriate use and interpretation}.
\newblock \bibinfo{journal}{\emph{Anesthesia \& analgesia}} \bibinfo{volume}{126}, \bibinfo{number}{5} (\bibinfo{year}{2018}), \bibinfo{pages}{1763--1768}.
\newblock


\bibitem[\protect\citeauthoryear{Strandberg, Himmelroos, and Gr{\"o}nlund}{Strandberg et~al\mbox{.}}{2019}]%
        {strandberg2019discussions}
\bibfield{author}{\bibinfo{person}{Kim Strandberg}, \bibinfo{person}{Staffan Himmelroos}, {and} \bibinfo{person}{Kimmo Gr{\"o}nlund}.} \bibinfo{year}{2019}\natexlab{}.
\newblock \showarticletitle{Do discussions in like-minded groups necessarily lead to more extreme opinions? Deliberative democracy and group polarization}.
\newblock \bibinfo{journal}{\emph{International Political Science Review}} \bibinfo{volume}{40}, \bibinfo{number}{1} (\bibinfo{year}{2019}), \bibinfo{pages}{41--57}.
\newblock


\bibitem[\protect\citeauthoryear{Wang, Gu, and Wang}{Wang et~al\mbox{.}}{2019}]%
        {wang2019causes}
\bibfield{author}{\bibinfo{person}{Yu-Huan Wang}, \bibinfo{person}{Tian-Jun Gu}, {and} \bibinfo{person}{Shyang-Yuh Wang}.} \bibinfo{year}{2019}\natexlab{}.
\newblock \showarticletitle{Causes and characteristics of short video platform internet community taking the TikTok short video application as an example}. In \bibinfo{booktitle}{\emph{2019 IEEE International Conference on Consumer Electronics-Taiwan (ICCE-TW)}}. IEEE, \bibinfo{pages}{1--2}.
\newblock


\bibitem[\protect\citeauthoryear{Welch}{Welch}{1967}]%
        {welch1967use}
\bibfield{author}{\bibinfo{person}{Peter Welch}.} \bibinfo{year}{1967}\natexlab{}.
\newblock \showarticletitle{The use of fast Fourier transform for the estimation of power spectra: a method based on time averaging over short, modified periodograms}.
\newblock \bibinfo{journal}{\emph{IEEE Transactions on audio and electroacoustics}} \bibinfo{volume}{15}, \bibinfo{number}{2} (\bibinfo{year}{1967}), \bibinfo{pages}{70--73}.
\newblock


\bibitem[\protect\citeauthoryear{White, Ruthven, and Jose}{White et~al\mbox{.}}{2002}]%
        {63white2002use}
\bibfield{author}{\bibinfo{person}{Ryen~W White}, \bibinfo{person}{Ian Ruthven}, {and} \bibinfo{person}{Joemon~M Jose}.} \bibinfo{year}{2002}\natexlab{}.
\newblock \showarticletitle{The use of implicit evidence for relevance feedback in web retrieval}. In \bibinfo{booktitle}{\emph{Advances in Information Retrieval: 24th BCS-IRSG European Colloquium on IR Research Glasgow, UK, March 25--27, 2002 Proceedings 24}}. Springer, \bibinfo{pages}{93--109}.
\newblock


\bibitem[\protect\citeauthoryear{Yang, Huang, Xia, Huang, Luo, and Lin}{Yang et~al\mbox{.}}{2023}]%
        {yang2023debiased}
\bibfield{author}{\bibinfo{person}{Yuhao Yang}, \bibinfo{person}{Chao Huang}, \bibinfo{person}{Lianghao Xia}, \bibinfo{person}{Chunzhen Huang}, \bibinfo{person}{Da Luo}, {and} \bibinfo{person}{Kangyi Lin}.} \bibinfo{year}{2023}\natexlab{}.
\newblock \showarticletitle{Debiased Contrastive Learning for Sequential Recommendation}. In \bibinfo{booktitle}{\emph{Proceedings of the ACM Web Conference 2023}}. \bibinfo{pages}{1063--1073}.
\newblock


\bibitem[\protect\citeauthoryear{Yarchi, Baden, and Kligler-Vilenchik}{Yarchi et~al\mbox{.}}{2021}]%
        {yarchi2021political}
\bibfield{author}{\bibinfo{person}{Moran Yarchi}, \bibinfo{person}{Christian Baden}, {and} \bibinfo{person}{Neta Kligler-Vilenchik}.} \bibinfo{year}{2021}\natexlab{}.
\newblock \showarticletitle{Political polarization on the digital sphere: A cross-platform, over-time analysis of interactional, positional, and affective polarization on social media}.
\newblock \bibinfo{journal}{\emph{Political Communication}} \bibinfo{volume}{38}, \bibinfo{number}{1-2} (\bibinfo{year}{2021}), \bibinfo{pages}{98--139}.
\newblock


\bibitem[\protect\citeauthoryear{Ye, Ai, and Liu}{Ye et~al\mbox{.}}{2024}]%
        {ye2024brain}
\bibfield{author}{\bibinfo{person}{Ziyi Ye}, \bibinfo{person}{Qingyao Ai}, {and} \bibinfo{person}{Yiqun Liu}.} \bibinfo{year}{2024}\natexlab{}.
\newblock \showarticletitle{Brain-Computer Interface Meets Information Retrieval: Perspective on Next-generation Information System}. In \bibinfo{booktitle}{\emph{Proceedings of the 1st International Workshop on Brain-Computer Interfaces (BCI) for Multimedia Understanding}}. \bibinfo{pages}{61--65}.
\newblock


\bibitem[\protect\citeauthoryear{Ye, Xie, Ai, Liu, Wang, Su, and Zhang}{Ye et~al\mbox{.}}{2023}]%
        {ye2023relevance}
\bibfield{author}{\bibinfo{person}{Ziyi Ye}, \bibinfo{person}{Xiaohui Xie}, \bibinfo{person}{Qingyao Ai}, \bibinfo{person}{Yiqun Liu}, \bibinfo{person}{Zhihong Wang}, \bibinfo{person}{Weihang Su}, {and} \bibinfo{person}{Min Zhang}.} \bibinfo{year}{2023}\natexlab{}.
\newblock \showarticletitle{Relevance Feedback with Brain Signals}.
\newblock \bibinfo{journal}{\emph{ACM Transactions on Information Systems}} (\bibinfo{year}{2023}).
\newblock


\bibitem[\protect\citeauthoryear{Ye, Xie, Liu, Wang, Li, Li, Chen, Zhang, and Ma}{Ye et~al\mbox{.}}{2022}]%
        {ye2022don}
\bibfield{author}{\bibinfo{person}{Ziyi Ye}, \bibinfo{person}{Xiaohui Xie}, \bibinfo{person}{Yiqun Liu}, \bibinfo{person}{Zhihong Wang}, \bibinfo{person}{Xuancheng Li}, \bibinfo{person}{Jiaji Li}, \bibinfo{person}{Xuesong Chen}, \bibinfo{person}{Min Zhang}, {and} \bibinfo{person}{Shaoping Ma}.} \bibinfo{year}{2022}\natexlab{}.
\newblock \showarticletitle{Why Don't You Click: Understanding Non-Click Results in Web Search with Brain Signals}. In \bibinfo{booktitle}{\emph{Proceedings of the 45th International ACM SIGIR Conference on Research and Development in Information Retrieval}}. \bibinfo{pages}{633--645}.
\newblock


\bibitem[\protect\citeauthoryear{Zheng, Liu, Lu, Lu, and Cichocki}{Zheng et~al\mbox{.}}{2018}]%
        {zheng2018emotionmeter}
\bibfield{author}{\bibinfo{person}{Wei-Long Zheng}, \bibinfo{person}{Wei Liu}, \bibinfo{person}{Yifei Lu}, \bibinfo{person}{Bao-Liang Lu}, {and} \bibinfo{person}{Andrzej Cichocki}.} \bibinfo{year}{2018}\natexlab{}.
\newblock \showarticletitle{Emotionmeter: A multimodal framework for recognizing human emotions}.
\newblock \bibinfo{journal}{\emph{IEEE transactions on cybernetics}} \bibinfo{volume}{49}, \bibinfo{number}{3} (\bibinfo{year}{2018}), \bibinfo{pages}{1110--1122}.
\newblock


\end{thebibliography}

\end{document}